\documentclass[useAMS,usenatbib]{mn2e}
\usepackage[applemac]{inputenc}
\usepackage[pdftex]{color,graphicx}
\usepackage{amsmath}
\usepackage{amssymb}
\usepackage{epsfig}
\usepackage[usenames,dvipsnames]{xcolor}
\usepackage{pst-node,xcolor,graphicx}
\usepackage{rotating}
\usepackage{hyperref}	
\def\kms{\mbox{km~s$^{-1}$}}
\def\kpc{\mbox{kpc}}

\renewcommand{\AA}{\protect\hbox{$\overset{_{\circ}}{\text{A}}$}}
\newcommand{\Ha}{H$\alpha$}

\newcommand{\hi}{H{\sc I}\, }

\def\aj{AJ}                 
\def\apj{ApJ}                 
\def\apjl{ApJL}                 
\def\apjs{ApJS}               
\def\mnras{MNRAS}             
\def\aap{A\&A}                 
\def\pasp{PASP}		
\def\nat{NAT}
\def\apss{APSS}

\title[Corrugated velocity patterns in the spiral galaxies: NGC 278, NGC 1058, NGC 2500 \& UGC 3574]
{Corrugated velocity patterns in the spiral galaxies: NGC 278, NGC 1058, NGC 2500 \& UGC 3574}
\author[M.C. S\'anchez-Gil et al.]{M. Carmen S\'anchez-Gil$^{1}$\thanks{E-mail: mcarmen.sanchez@uca.es}, 
Emilio J. Alfaro$^{2}$ and Enrique P\'erez$^{2}$.\\
$^{1}$Universidad de C\'adiz, Facultad de Ciencias, Puerto Real, Spain\\
$^{2}$Instituto de Astrof\'{i}sica de Andaluc\'ia (CSIC), E18008, Granada, Spain} 

\begin{document}

\date{Accepted 2015 September 22. Received 2015 May; in original form 2015 May 9}

\pagerange{\pageref{firstpage}--\pageref{lastpage}} \pubyear{2015}

\maketitle

\label{firstpage}

\begin{abstract}
We address the study of the \Ha\ vertical velocity field in a sample of four nearly face-on galaxies using long slit spectroscopy taken with the ISIS spectrograph attached to the WHT at the Roque de los Muchachos Observatory (Spain). The spatial structure of the velocity vertical component shows a radial corrugated pattern with spatial scales higher or within the order of { one} kiloparsec. The gas is mainly ionized by high-energy photons: only in some locations of NGC~278 and NGC~1058 is there some evidence of ionization by low-velocity shocks, which, in the case of NGC~278, could be due to minor mergers. The behaviour of the gas in the neighbourhood of the spiral arms fits, in the majority of the observed cases, with that predicted by the so-called hydraulic bore mechanism, where a thick magnetized disk encounters a spiral density perturbation. The results obtained show that it is { difficult to explain the \Ha\ large scale velocity field without the presence of a magnetized, thick galactic disk}. Larger samples and spatial covering of the galaxy disks are needed to provide further insight into this problem.
\end{abstract}


\section{Introduction}

The Milky Way shows a galactic disk quite far from the planarity \citep{1957Natur.180..677K}. This high degree of structure was first observed in the gas and can be classified into three categories associated with different regions and spatial scales of the disk: 
a) the gas in the innermost region of the galaxy appears to be tilted with respect to the formal galactic plane \citep{1960MNRAS.121..132G}{ ; b)} warping of the gas in the outer regions of the disk is almost a ubiquitous feature of spiral galaxies in the Local universe, even for those  that are seemingly isolated \citep[i.e.][]{1960MNRAS.121..132G,1973A&A....27..407V,1991MNRAS.251..193F}{ ;} and c) since the 1950s we { have known} that \hi in the Milky Way shows a coherent wavy structure with respect to the fundamental plane of the Galaxy in the inner region of the Sun called ``corrugations" \citep[i.e.][]{1977AJ.....82..408L,A96}.

Corrugations have also been observed in different stellar { populations} of the Milky Way \citep[i.e.][]{1977Ap&SS..50..281Q,1992ApJ...399..576A}
 and both in the azimuthal direction along different spiral arms \citep{1986A&A...163...43S}, { and} in the radial direction at different galactocentric radii  \citep{1995fmw..conf...87M}.  
This { phenomenon} is not just limited to our Galaxy but { has} also been observed in external galaxies \citep[e.g.][]{1991MNRAS.251..193F,2008ApJ...688..237M}, and for different tracers of both gas and stellar  populations. 
In addition, although most of the stellar corrugations were determined from the distribution of young stars, infrared observations seem to show that old stellar { populations} also { participate} in this structure \citep{1989ApJ...341L..13D,1995fmw..conf...91R}.

All these results suggest that this is a universal phenomenon common in late-type galaxies, observable at almost any wavelength \citep{A96,2008ApJ...688..237M}. However, despite the universality of  this phenomenon it has { barely} been studied and { is} poorly understood.

These spatial corrugations clearly must be associated with wavy vertical motions in the galactic plane. Evidence of these kinematic waves was first detected in the analysis of the rotation curves of spiral galaxies \citep[e.g.][]{1963ApJ...137..363D,1965BOTT....4....8P}, but it was not until \citet{2001ApJ...550..253A} analysed the velocity corrugations in NGC~5427  in more detail that a likely physical mechanism was proposed for their origin.

Henceforth, we use the term ``velocity corrugation" to refer to the observed wavy field of the velocity vertical component. The two kinds of structures have been observed in spiral galaxies: a) spatial corrugations in our Galaxy and other edge-on external galaxies, and b) velocity corrugations in face-on galaxies. 

Given the nature of the problem, galaxy-observer geometry plays a fundamental role in the interpretation of the observables. Within our Galaxy we can easily observe the corrugated structure of the spiral arms in the solar neighbourhood in the optical range, as well as the HI structure at larger distances in the radio wavelength range. 
The vertical velocity field of these morphologies shows an almost null component in the direction of the line of sight. 
However, \citet{1985IAUS..106..175F} were able to detect true velocity gradients perpendicular to the Galactic plane that  could not be totally explained by geometric effects.  

By contrast, the face-on external galaxies represent the best natural laboratory for the study of the vertical velocity field, given that most of the observed radial velocity is representative of the galaxy disk velocity vertical component. However, they do not provide any information about spatial corrugations. 

In summary, in the last 60 years several  pieces of evidence have been collected showing that most disks of spiral galaxies present  a residual wavy structure with respect to the galactic main plane. 
This undulatory  behaviour is observable at different wavelengths, being representative { of the} gas in different physical conditions, as well as of different stellar populations. Different scenarios have been proposed to explain this phenomenon \citep[see][]{A96,2004ASSL..315.....A}, but there is a lack of systematic studies 
which would help to better understand these structures and their possible origin.  

In this paper, we address the study of the vertical velocity in a sample of four nearly face-on spiral galaxies. As aforementioned, the geometry of the problem is a main criterion in the selection of the sample as well as of the long slit azimuthal position angles. 
In Section \ref{SecSample} we show the galaxy sample and their selection criteria. Observations and data reduction details are presented in Section \ref{SecObs}. 
In section \ref{SecVz} we describe the geometry of the problem and establish its fundamental  relationships.
The main results and relationships between the vertical component of the velocity field and the \Ha\ emission lines are shown in Section \ref{SecResults}. In Section \ref{SecDD} we present an analysis of { the ionization sources via Diagnostic Diagrams}. Finally, a summary of this work can be found in Section \ref{SecSum}.

\section{Galaxy Sample}\label{SecSample}

\begin{figure*}
\setlength{\unitlength}{1cm}
\begin{picture}(18,16)
\put(0,8){\includegraphics[width=8cm]{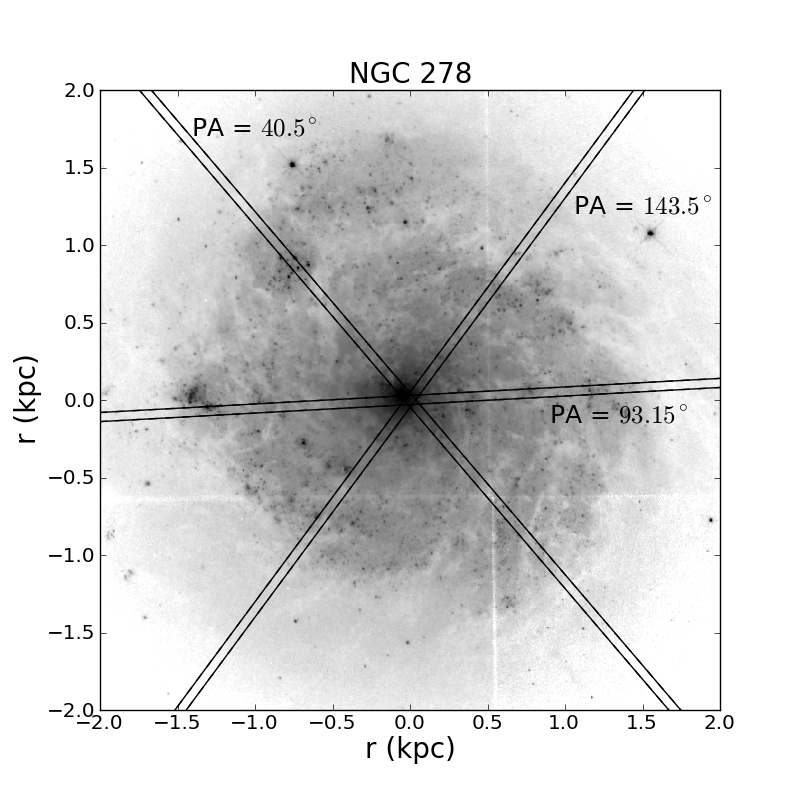}}
\put(8,8){\includegraphics[width=8cm]{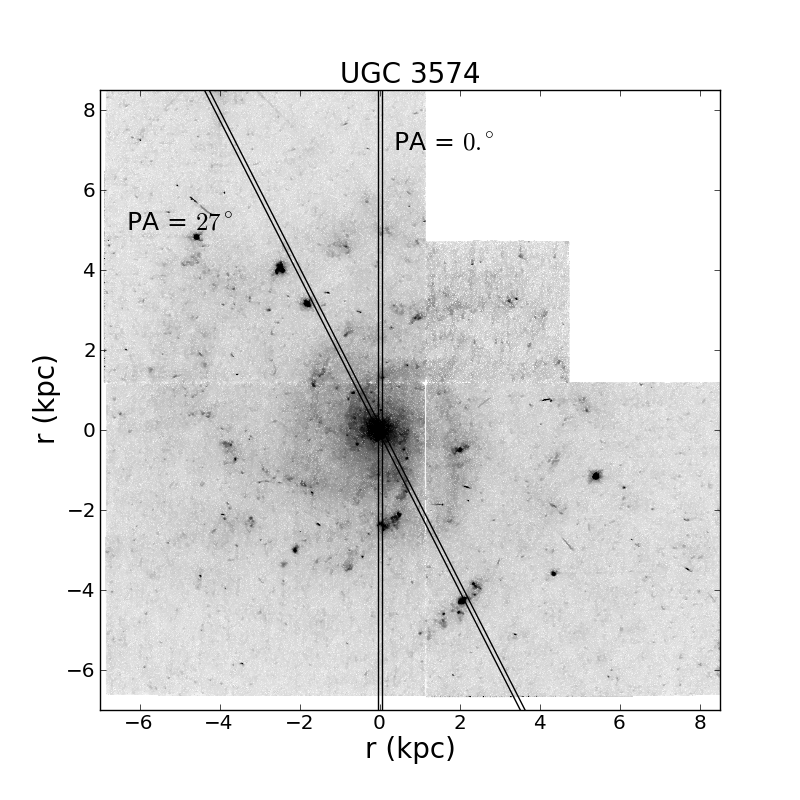}}
\put(0,0){\includegraphics[width=8cm]{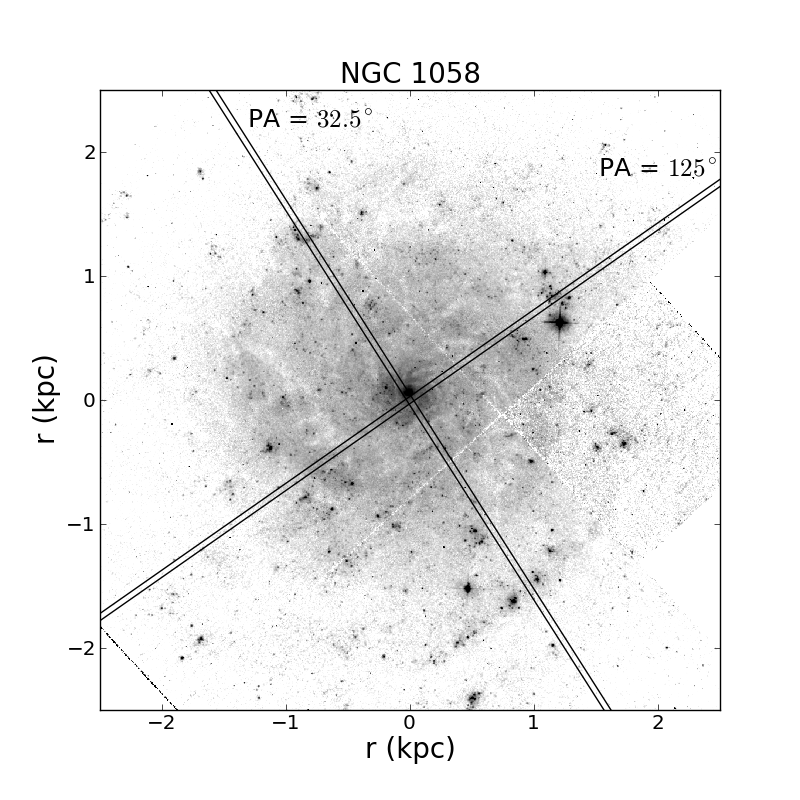}}
\put(7.75,0.){\includegraphics[width=10cm]{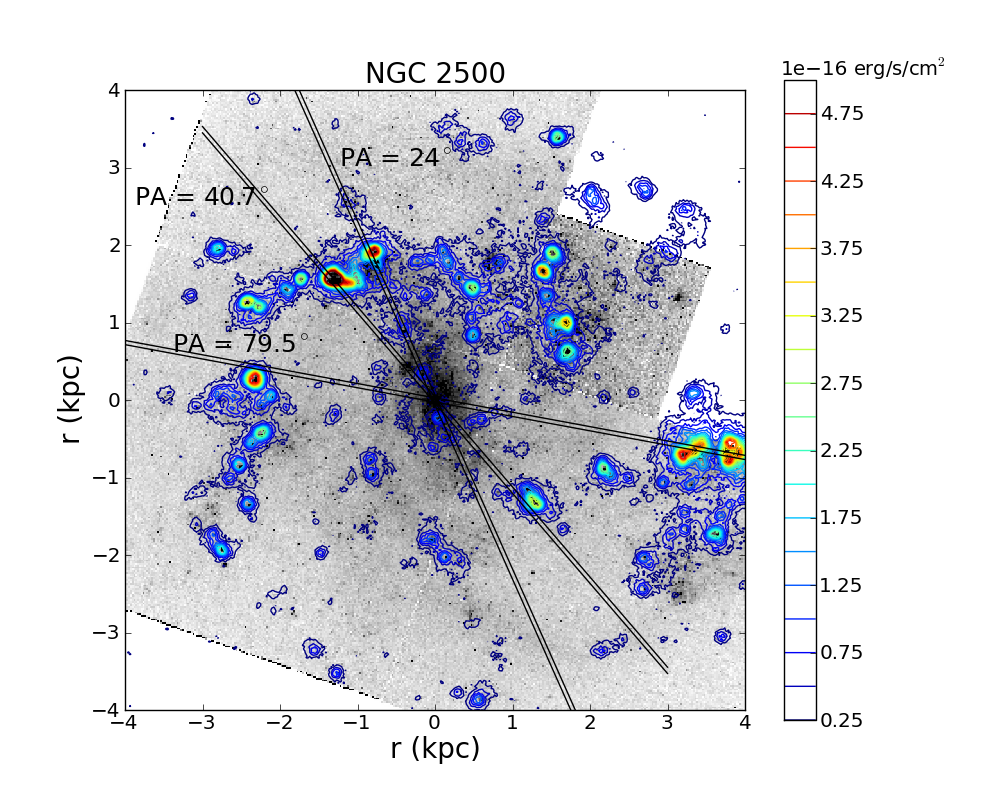}}
\end{picture}
\caption{ Images for the galaxies of the sample: NGC~278, NGC~1058, NGC~2500 and UGC~3574 ( HST images, WFPC2 instrument and F450W filter). { North is top and East to the left}. The different slit positions of the long-slit spectra, taken with the red arm ISIS spectrograph attached to the 4.2m WHT, are also represented on the image{ .} 
In the case of NGC~2500 an \Ha\ image { is also added}, taken from the LVL survey, plotted  as { coloured} iso-contour lines onto the HST image. The contours { range} from 2.5E-17 to 5.E-16 erg/s/cm$^2$.}
\label{fig1}
\end{figure*}

The galaxies were selected from the LEDA database\footnote{Lyon-Meudon Extragalactic Database, http://leda.univ-lyon1.fr/} with the  criteria: 
(i) low inclination angle ( i$<25^{\circ}$)\footnote{For NGC~1058, LEDA gives $i=58.5^{\circ}$, but its correct inclination angle is in the range  $4^{\circ}-11^{\circ}$ (Lewis 1987; van der Kruit \& Shostak 1984; 2007AJ....134.1952P)}; 
(ii) nearby ($v<4000$ \kms); (iii) a { well-defined} spiral morphological type (2$<$t$<$8); and (iv) a diameter larger than 2 arcmin (D$_{25}>2$). 

This information and other basic properties of the galaxy sample are summarized in Table \ref{tab1}.  
With the aim of having a uniform and homogeneous criterion for the inclination and for the position angle, and since these values can change significantly  between different authors, we show the values given by \citet{2008MNRAS.390..466E}.
We also take the data for the calculation of the rotation curves  from these authors. 

Figure \ref{fig1} displays  images of the galaxies with the different slit positions. Next, we summarize the main references and properties for these galaxies found in the literature. 

\begin{itemize}

\item {\it NGC 278} 
is  isolated \citep{1988ApJ...335...74B} and almost face-on, 
part of the local supercluster of galaxies centred around the Virgo cluster \citep{1993MNRAS.264..665V}.  
It is relatively small in the optical emission, with size of about $2'$ (Israel 2009) corresponding to a diameter of 7 kpc 
at a distance of $\sim12.1$ Mpc (see Table \ref{tab1}). 
However, in HI its diameter is five times larger \citep{2004A&A...423..481K}.

In optical images its morphology is that of a small, non-barred galaxy with a bright star-forming nucleus and spiral arms. 
Its nucleus is not only quite bright in H$\alpha$, but also in the infrared continuum \citep{2000AJ....120..583D,2003JKAS...36..149S},
and in the [CII] line \citep{2001ApJ...561..766M}, consistent with intense star-forming activity \citep{1995ApJS...98..477H,2002ApJS..143...73E}.
There is no indication of a low-luminosity AGN \citep{2007A&A...468..129T}.  

NGC~278 can be found in many recent works, but mostly forming part of  
surveys \citep[e.g.][]{2009A&A...506..689I}. A specific study of this galaxy can be found in
\citet{2004A&A...423..481K}, who concluded that its appearance most likely
reflects a recent minor merger. 
Asymmetries set up in the disk by this minor merger may well be the cause of the intense star formation in the inner region, which can be interpreted as a rare example of a nuclear
ring in a non-barred galaxy. Rather than being induced by the secular evolution of a bar, 
this nuclear ring would then be the direct result of an interaction event in the recent history of the galaxy.
It also shows a disturbed velocity field in its \Ha\ kinematics, which may be partly the result of spiral
density wave streaming motions. 
The \hi emission, although faint, also shows disturbed morphology and kinematics.   

\item {\it NGC~1058} 
is a nearly face-on spiral galaxy with an inclination angle in the range $4^{\circ}$ to $11^{\circ}$ \citep{2007AJ....134.1952P}. 
The distance to this galaxy is uncertain, ranging from 8.4 Mpc \citep{2000A&A...357..437B} to 14.5 Mpc \citep{1974ApJ...194..559S}. 
We adopt a distance of 9.8 Mpc, given by a more recent work \citep{2006ApJS..164...81M}, for which 1\arcsec\ corresponds to 47.5 pc.

It has a clear knotty structure not organized into spiral arms \citep{2000A&A...357..437B}, with a complete absence of any well-defined arms in \Ha\ emission, and a flocculent appearance of the  disk's innermost regions \citep{1995IAUS..164..425F}.
These features are also observed in the HST image (Figure \ref{fig1}). 

Radial metallicity and reddening gradients, both decreasing with the galactocentric distance, are found from studies of \Ha\ emitting regions by \citet{1998ApJ...506L..19F}, a trend often observed in spirals \citep{2014&A...563A..49S}. 

The \hi velocity dispersion (ranging  4 -- 15 \kms) also decreases with radius, 
but the decline of starlight with radius is much steeper than that of the velocity dispersion \citep{2007AJ....134.1952P}. 
These authors also find  that the \hi velocity dispersion is not correlated with star formation or the spiral arms, so a global radial falloff must be explained in the context of significant local effects. 

\citet{2000A&A...357..437B} also found a steeper radial slope for the star formation rate (SFR) in NGC~1058, with young star associations relatively small (peak at $\sim$ 50 pc), the star formation being clearly stronger in the central regions.
However, from our long slit spectra, Figure \ref{fig1},  \Ha\  reveals stronger emission in the outer than in the innermost regions.

\item {\it UGC~3574} 
is a nearby late-type spiral galaxy, with low inclination; at a distance of 21.8 Mpc 
\citep{2004A&A...414...23J} it is the furthest galaxy of the sample. 
Two faint spiral arms can be distinguished, while the stronger emission comes from the nuclear region and some very compact and clumpy sources at the end of the southern spiral arm or across the northern one. 

No specific studies for this galaxy are found in the literature, but it is included in many surveys for different purposes, e.g. \citet{2002AJ....123.1389B} who study the nuclear star clusters in late-type spiral galaxies. 
In this work, UGC~3574 is characterized as having a ``naked" nuclear cluster in the I-band WFPC2 image, forming a distinct and isolated entity within the disk and without any signs of  a kinematic centre.

\item {\it NGC~2500} 
belongs to a quartet of galaxies \citep{1994cag..book.....S}. \citet{2008MNRAS.390..466E} find diffuse \Ha\ emission in their observed \Ha\ maps, in agreement with \citet{2004A&A...414...23J}, where its short bar is almost aligned with its minor kinematical axis. 
There are no individual works for this galaxy, mainly found as part of surveys, e.g. 
\citet{2008MNRAS.390..466E} or \citet{2004A&A...414...23J}. 

In the HST/WFPC2+F450W image, Figure \ref{fig1}, it is observed that the star formation is mainly concentrated in the nuclear region and its  well-defined, northern spiral arm. The southern arm is fainter and, globally, the southern region of this galaxy appears more flocculent.  

These features are enhanced in \Ha\ emission{, as } can be observed in Figure \ref{fig1}. The \Ha\ image is taken from the public database of the Local Volume Legacy (LVL) survey\footnote{http://irsa.ipac.caltech.edu/data/SPITZER/LVL/}, and is over-plotted to the HST image, in flux scaled contours.  

\end{itemize}

\section{Observations and data reduction}\label{SecObs}

We obtained long-slit spectroscopy with the double arm ISIS spectrograph attached to the  4.2m William Herschel Telescope (WHT), at the Roque de los Muchachos Observatory (La Palma) during December 2003. 
This instrument consists of two intermediate dispersion spectrographs operating simultaneously, separating the blue and red spectral regions, which are imaged onto an EEV12 and a  MARCONI2 CCD chip, respectively. 
In this way we have two spectral ranges observed simultaneously, a blue one  centred around H$\beta$ (4861 \AA), and a red one around \Ha\ (6563 \AA).  
The gratings used, R1200R  and R600B, provide a dispersion of 0.23 and 0.45 \AA/pixel respectively. 
The slit width of 1 arcsec projects onto about 3.64 pixels Full-Width-Half-Maximum (FWHM) on the detector; the spatial sampling along the slit is 0.2 arcsec/pixel. 
The slit was placed at two or three different position angles for each galaxy (Table \ref{tab2}). 
An angle close to the minor axis of the galaxy was chosen, so the projection of the rotation velocity is nearly negligible and the observed velocities are mainly the projected vertical component on the line of sight. Moreover, other intermediate angles were taken to compare corrugated patterns with the latter, and/or to cross some brighter \Ha\ features. 

The spectra were reduced and calibrated following the standard procedure. Bias subtraction, flat-fielding, wavelength and flux calibration were done with the IRAF\footnote{IRAF is distributed by the National Optical Astronomy Observatory, which is operated by the Association of Universities for Research in Astronomy (AURA) under cooperative agreement with the National Science Foundation} task {\tt ccdproc}. For the wavelength calibration CuNe $+$ CUAR lamps were used. 
The standard stars Feige 34, g191b2b, gd248, gd50 and hz21 from the Oke (1990) catalogue were used for flux calibration. 
Sky subtraction was done using the IRAF NOAO package task {\tt background}. 
For each wavelength range, the different exposures at each slit orientation were combined, eliminating cosmic rays and bad pixels, thus obtaining the final spectra. 

The observed velocity along the slit is obtained from the \Ha\ emission line fitting. Moreover, 
to calculate the diagnostic diagrams to study the ionization (section \ref{SecDD}), 
we  fit the emission lines \Ha, [NII], [SII], H$\beta$, and [OIII], with a Gaussian at each pixel along the slit where the emission was strong enough to be fitted. Line flux, FWHM, {\Large }central wavelength, and their corresponding errors, are calculated with the STARLINK\footnote{http://star-www.rl.ac.uk/star/docs/sun50.htx/sun50.html} package DIPSO. The continuum level is simultaneously fitted locally with a first order polynomial.

\begin{table*}
\centering
\caption{Galaxy Parameters$^a$}
\begin{tabular}{@{}lccccccccc@{}}
\hline
Galaxy & RA (J2000) & Dec. (J2000) & Type & Redshift & Dist.$^b$\ & Inclin.$^c$ & PA$^c$& Dimensions & M$_B$ \\
 & h~m~s & $^\circ$~$'$~$''$ & &    & (Mpc)  & (deg)&(deg)&(arcmin) &   \\
\hline
NGC~278 & 00 52 04.3 & +47 33 02 & SAB(rs)b & 0.002090 & 12.1&21$\pm$14&52$\pm$ 3 & 2.1$\times$2.0 & -19.6 \\
NGC~1058 & 02 43 30.0 &+37 20 29 &SA(rs)c &0.001728 & 9.8 & 6 $\pm$ 15&125$\pm$ 6& 3.0$\times$2.8& -18.7 \\
UGC~3574 &06 53 10.4 &+57 10 40 &SA(s)cd &0.004807 & 21.8 & 19 $\pm$ 10&99$\pm$ 3& 4.2$\times$3.6&-18.0 \\
NGC~2500 & 08 01 53.2 &+50 44 14 &SAB(rs)d &0.001715 &11.0 & 41$\pm$ 10&85$\pm$ 2&2.9$\times$2.6 &-18.2 \\
\hline
\label{tab1}
\end{tabular}
\begin{tabular}{rl}
$^{a}$ & Sourced from {\it NASA Extragalactic Database} \\
$^{b}$ & Moustakas \& Kennicutt 2006; except for UGC~3574, James et al. 2004\\
$^{c}$ & Epinat et al. 2008. Inclination and position angle deduced from their
analysis of the velocity field (from \Ha\ data). \\
& { The galaxies were selected from the LEDA database$^1$ with a criterium of a low inclination angle ( i$<25^{\circ}$). NGC~2500 }\\
& {  appears in this catalog with an inclination angle of 18$^{\circ}$, but we use data from Epinat et al. (2008) to determine the galactic }\\
& { rotation curve, where a kinematic inclination angle  of 41$^{\circ}$ is estimated. For the sake of coherence in our data, we take this } \\
&{ value into our analysis.}
\end{tabular}
\caption{Journal of observations}
\begin{tabular}{@{}l c c c c c c@{}}
\hline
Galaxy&Date & P.A. (deg) & Wavelength range (\AA) &Exposure time (s) \quad & Slit (\arcsec)& Airmass\\ 
 \hline
NGC~278& 15.12.2003& 40.5, 93.15, 143.5& 6034$-$7088&3$\times$1200 & 1.03&1.06 - 1.33\\
NGC~2500 & 15.12.2003&24, 79.5, 40.7& 6034$-$7088& 2,3$\times$1200& 1.03&1.08 - 1.14\\
NGC~1058 & 16.12.2003&32.5, 125& 6034$-$7088& 3$\times$1800& 1.03&1.01 - 1.17\\
UGC~3574 & 16.12.2003&0, 27, 122& 6034$-$7088& 3$\times$1200; 2,3$\times$1800& 1.03&1.14 - 1.28\\
\hline
\end{tabular}
\label{tab2}
\end{table*}%

\section{V$_Z$ calculation}\label{SecVz}

In the case of face-on galaxies, the vertical component of the disk velocity  field may be obtained  straightforwardly. As our  galaxies are nearly face-on, the observed velocity is the result of the sum of several projected velocity components, due to their inclination angle. This sum can be summarized as the vertical component to the disk, $ V_{\perp}$, plus its parallel component $V_{\parallel}$. 
We assume then that the observed velocity can be expressed in terms of 

\begin{eqnarray}
V_{obs} \mskip-12mu &=&\mskip-12mu V_{sys} + [V_{\parallel}  \sin i + V_{\perp} \cos i ] = \nonumber \\
 \mskip-12mu &=&\mskip-12mu V_{sys} + [(V_{rot} \cos \theta + V_{exp} \sin \theta) \sin i + V_Z \cos i ]
\label{eqVOBS}
\end{eqnarray}
where V$_{sys}$ is the systemic velocity of the galaxy (radial velocity of the galactic { centre} with respect to LSR), V$_{rot}$ and V$_{exp}$ the rotational and expansion velocity respectively, at the plane of the disk, and V$_Z$ is the velocity perpendicular to the galaxy main  plane; 
$\theta$ is the angle in the plane of the galaxy (counterclockwise from the major axis), and $i$ is the inclination angle of the galaxy disk.

We consider the x-axis in the direction of the maximum diameter passing through the galaxy centre
and the north equatorial pole, contained in the plane of the sky. And  the x-axis coincides with the major axis in the plane of the galactic disk.
Thus, the polar coordinates in the plane of the sky, (r,$\phi$), and in the plane of the galaxy, (R,$\theta$), are related as follows:

\begin{eqnarray}
R \mskip-12mu &=&\mskip-12mu r \sqrt{\cos^2(\phi-PA) + \frac{\displaystyle\sin^2(\phi-PA)}{\displaystyle cos^{2} i}} \label{eqrR}\\
R\cos\theta \mskip-12mu &=&\mskip-12mu r\cos(\phi-PA)\label{eqxX} \\
R\sin\theta \mskip-12mu &=&\mskip-12mu { r\sin(\phi-PA)}{ cos^{-1} i }\label{eqyY}
\end{eqnarray}

From equation \ref{eqVOBS}, neglecting the expansion velocity (i.e., assuming $V_{exp} = 0$), and applying the relations between the celestial coordinates (equations \ref{eqrR} and \ref{eqxX}),  the vertical velocity component can be  calculated as:

\begin{eqnarray}
V_{Z} \mskip-12mu &=&\mskip-12mu {\left( V_{obs}  - V_{sys} - V_{rot} \frac{r}{R}\cos(\phi - PA) \sin i  \right)}{\cos^{-1} i }= \nonumber \\
\mskip-12mu &=&\mskip-12mu  \left( V_{obs}  - V_{sys} - V_{rot}\frac{\sin i}{\sqrt{1+\frac{\displaystyle\tan^2(\phi - PA)}{\displaystyle\cos^{2} i}}}  \right) \cos^{-1} i 
\nonumber \\
\nonumber \\
\mskip-12mu &=&\mskip-12mu ( V_{obs}  - V_{sys})\cos^{-1} i \, - \,  V_{rot} \cos\theta \tan i 
\label{eqVZ}
\end{eqnarray}
where (from equations \ref{eqrR} and \ref{eqxX})

\begin{eqnarray}
\cos\theta \mskip-12mu &=&\mskip-12mu \frac{r}{R}\cos(\phi - PA) = \nonumber \\
\mskip-12mu &=&\mskip-12mu  \frac{\cos(\phi - PA)}{\sqrt{\cos^2(\phi-PA) + \sin^2(\phi-PA)/\cos^{2} i}} = \nonumber \\
\mskip-12mu &=&\mskip-12mu \left( 1 + \tan^2(\phi-PA)/\cos^{2} i \right)^{-1/2}
\label{eq_Acostheta}
\end{eqnarray}

The vertical velocities are  therefore computed as a function of V$_{sys}$, the systemic velocity of the Galaxy taken from the literature; $V_{obs}$ and $\phi$, the observed velocity, obtained by fitting the \Ha\ emission lines from our long-slit spectra, and their corresponding slit position angles; and  V$_{rot}$, $PA$ and $i$, the rotational velocity of the galaxy, the position angle of the line of nodes and the inclination angle of the galactic disk.  

\begin{table*}
\caption{Parameters from the rotation curve fit for each galaxy.}
\begin{center}
\begin{tabular}{ccccc}
\hline
Galaxy & $V_0$ & $r_{pe}$&$\alpha$& R \\
 & (km s$^{-1}$) & (arcsec) & & (correlation)\\
 \hline
  NGC~278 & 219.73$\pm$59.1& 1.08$\pm$0.36&-0.40$\pm$0.17 & 0.907\\
  NGC~1058 & 74.5$\pm$12.7 & 0.16$\pm$0.10 & 0.05$\pm$0.03& 0.668\\
  UGC~3574 & 155.5$\pm$18.5 & 1.29$\pm$0.31 & 0.024$\pm$0.017 & 0.916\\
  NGC~2500 & 55.9$\pm$3 & 0.40$\pm$0.53 & 0.04$\pm$0.007 & 0.872\\
\hline
\end{tabular}
\end{center}
\begin{flushleft}
** The rotation curve has been fitted by a least squares fit, following the parametric model of \citet{2002ApJ...571L.107G}: 
$V_{pe}(r) = V_0 (1-e^{-r/r_{pe}}) (1+\alpha \, r/r_{pe})$.
\end{flushleft}
\label{tab3}
\end{table*}%

As a first approximation, we consider a global velocity model where the main movement in the galactic plane is the rotational velocity, neglecting any other kind of kinematics, and it is assumed that they depend on the galactocentric distance or radius, $v_{rot}(R)$. We try to explain those movements that do not fit to the model as vertical motions { with} respect to the galactic plane. 

The observed velocities $V_{obs}$ are obtained from the \Ha\ emission line fitting of the long-slit spectra, described in Section \ref{SecObs}  above. 

\begin{figure*}
\centering
\includegraphics[width=0.49\textwidth]{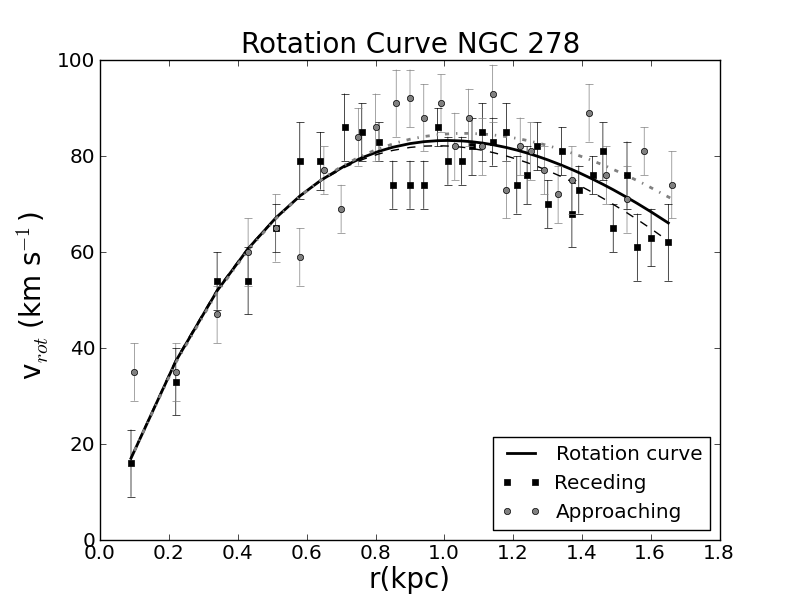}
\includegraphics[width=0.49\textwidth]{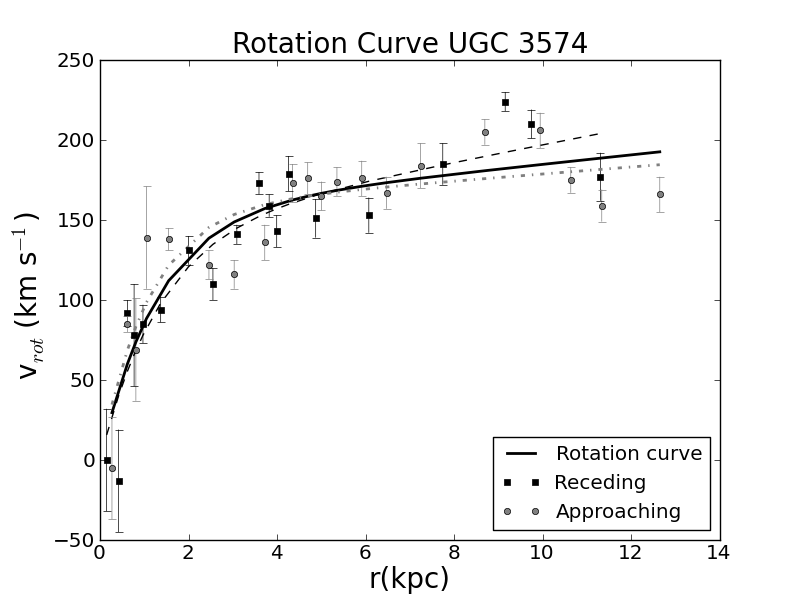}
\includegraphics[width=0.49\textwidth]{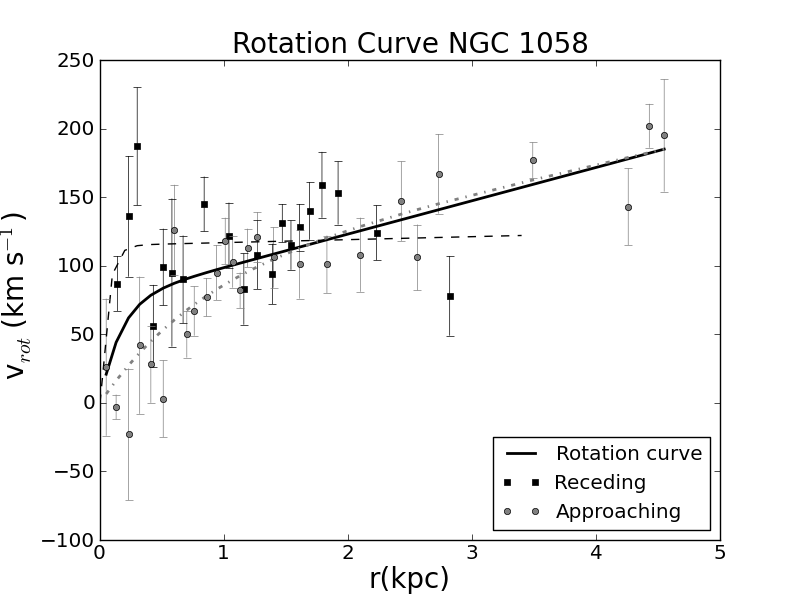}
\includegraphics[width=0.49\textwidth]{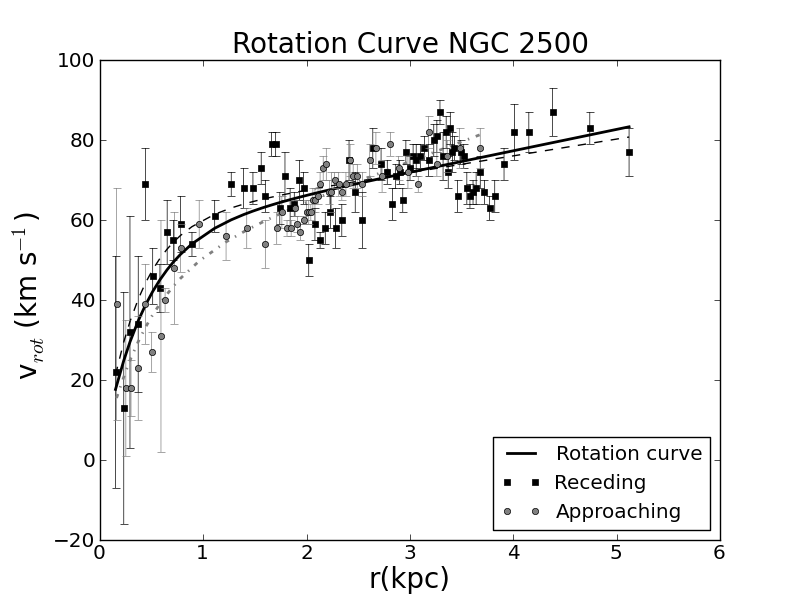}
\caption{Rotation curves for the galaxy sample. Data  are taken from \citet{2008MNRAS.390..466E}, where the approaching side of the curve is represented by grey circles, and the receding side by black squares. Both sides are fitted separately, represented with dotted and dashed lines respectively, and as a whole data set plotted with the bold line. This last fit is assumed as the rotation curve for the galaxy. Lopsidedness between the rotation curves of the receding and approaching sides is observed.}
\label{fig2}
\end{figure*}

Data for the calculation of the rotation curves $v_{rot}(R)$, inclination angles and PAs for each galaxy are taken from \citet{2008MNRAS.390..466E}, see Table \ref{tab1}. Rotation curves data are given as a set of points with their errors (Figure \ref{fig2}), so the rotation curves are obtained by fitting these sets of points using the parametric model of \citet{2002ApJ...571L.107G}, 

\begin{equation}
V_{pe}(r) = V_0 (1-e^{-r/r_{pe}}) (1+\alpha \, r/r_{pe})
\label{eqRotCurve}
\end{equation}
where $V_0$ regulates the overall amplitude of the rotation curve, $r_{pe}$ yields a scale length for the inner steep rise, and $\alpha$ sets the slope of the slowly changing outer part. After a least square fit of this model to the data, the parameters obtained for each galaxy are summarized in Table \ref{tab3}.

Figure \ref{fig2} shows the rotation curve fitted to the data of each galaxy. The approaching side of the curve is represented by  grey circles and the receding side by black squares. Both sides of the curve were fitted separately, represented with dashed and dotted lines, respectively.  The kinematic lopsidedness commonly observed in the dynamics of galaxies  can be clearly seen. This lopsidedness occurs in two different ways, as already described by \citet{2011A&A...530A..29V}, who  found five different types for the kinematic behaviour of disk galaxies based on the rotation curves of the receding and approaching sides separately. NGC~278 shows their third type, { in which} receding and approaching sides agree well at small radii but differ at large radii{, whereas the other three galaxies show type 5, in which} the curves of receding and approaching sides change sides at a certain radius. 

Despite this lopsidedness, the rotation curves were fitted to both receding and approaching points as a whole data set, given by the bold line in Figure \ref{fig2}. This global fit remains at intermediate values between the rotation curves of the receding and approaching sides, as a combination of both sides. 

Finally, as $V_Z$ is expressed as a function of $V_{obs}\pm\delta V_{obs}$ and $V_{rot}\pm\delta V_{rot}$ ( $i\pm\delta i$ and $PA\pm\delta PA$  are consider part of our data), we can calculate its uncertainty by propagating the error as:
 \begin{equation}
\delta V_Z = \lvert cos^{-1} i\rvert\delta V_{obs} + 
\lvert - \cos\theta \tan i \rvert\delta V_{rot} 
\label{eqErrorVz}
\end{equation}

\vspace{0.8cm}

\begin{figure*}
\includegraphics[width=0.45\textwidth]{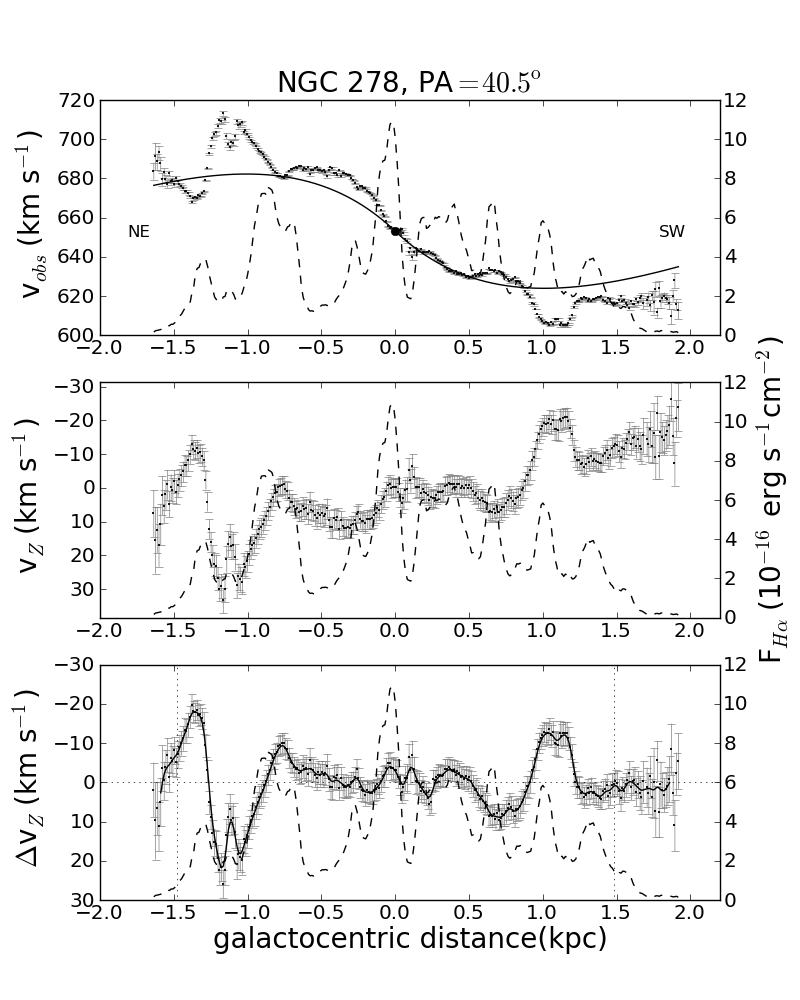}
\includegraphics[width=0.45\textwidth]{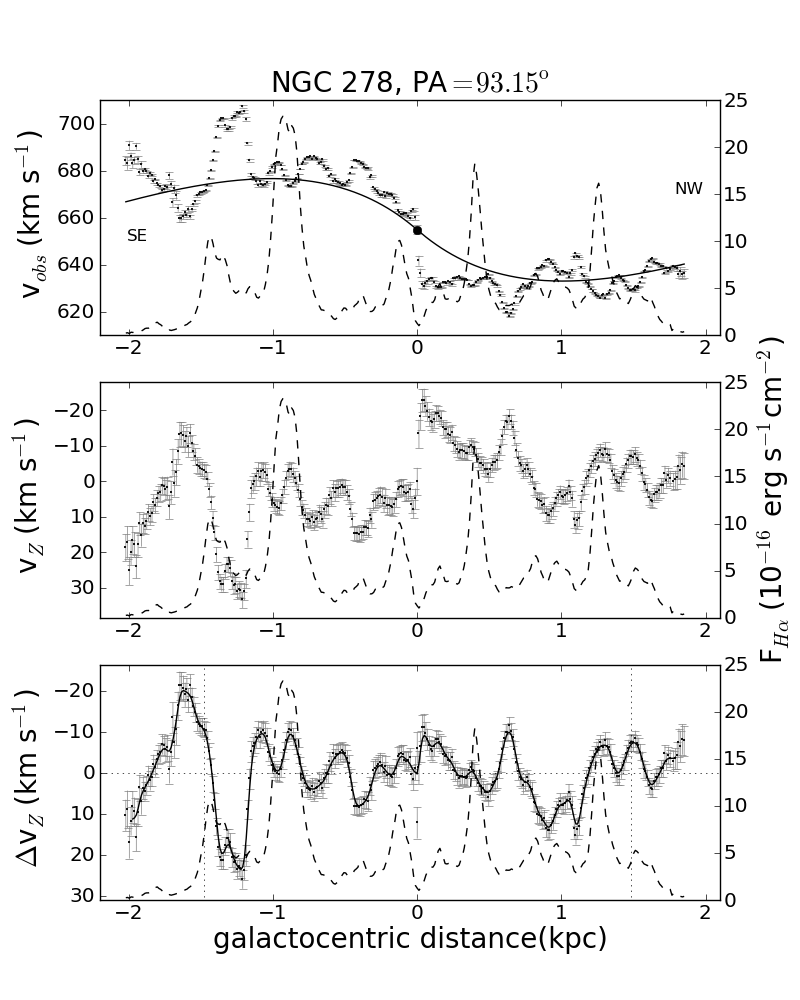}
\includegraphics[width=0.45\textwidth]{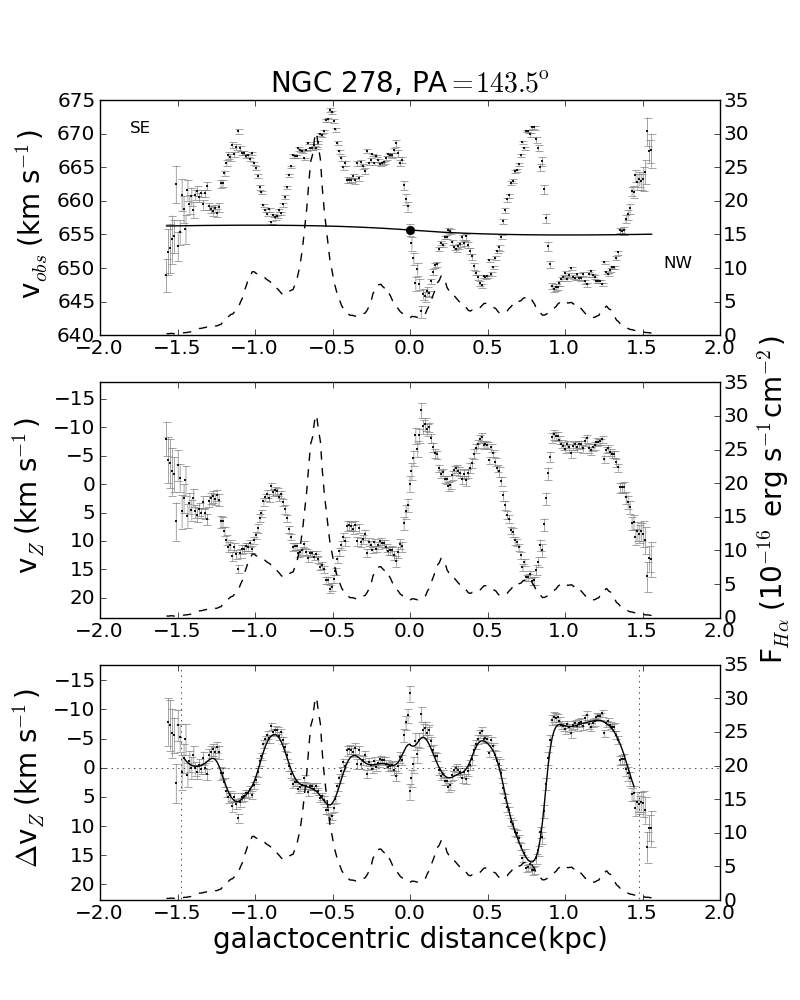}
\caption{  
Top: The observed velocities for NGC~278, derived from the \Ha\ emission lines (see Section \ref{SecObs}). 
The bold line corresponds  to the projected rotation curve, and the black dot marks the systemic velocity. 
The \Ha\ intensity is drawn in the right-hand ordinate, with a dashed line. 
The {\it x-axis} represents the distance to the reference pixel, in kpc units;
{ Centre}: The perpendicular velocity $V_Z$, at each pixel position across the slit, calculated as Equation \ref{eqVZ};
Bottom: The detrended perpendicular velocity, by fitting a linear component to the global trend.  
A dotted line marking the zero value  stands out as the global trend has been practically removed,  with only the local oscillations of $V_Z$ remaining.
The black line is a smoothing line of $\Delta V_Z$, for an easier comparison with the \Ha\ emission lines.}
\label{fig3}
\end{figure*}

Figure \ref{fig3}  shows the procedure  for obtaining the vertical motions for the galaxy NGC~278, at each of the three position angles. 
The top panels show the observed radial velocities $V_{obs}$ with their error bars, together with the projection of the rotation curve $V_{rot}$ (the bold line) fitted as described above. The systemic velocity $V_{sys}$ is marked by a  black-filled circle. 
All these plots have the galactocentric distance in the $x$-axis, where zero distance marks the galactic centre position. 
The \Ha\ emission line flux is also plotted, with the aim of looking for some relation or correlation between the velocity and flux peaks. 

The middle panels show that $V_Z$ has some residual trend, as residual of the rotational velocity removal probably due to the  described lopsidedness.  
To remove these rotational residues, the vertical velocity component has been detrended, hereafter denoted as $\Delta V_Z$, by fitting a linear component to the gentle rise for each rotation curve side, the approaching and the receding sides. 
In this simple way, the residual trend from the difference between the approaching and the receding sides of the rotation curve is removed; and thus only the local fluctuations, or oscillations, of the vertical  velocity component across the disk, throughout the spiral arms and for different slit positions{, remain}. 
The bottom panels show the detrended velocities $\Delta V_Z$.

\begin{figure*}
\includegraphics[width=0.45\textwidth]{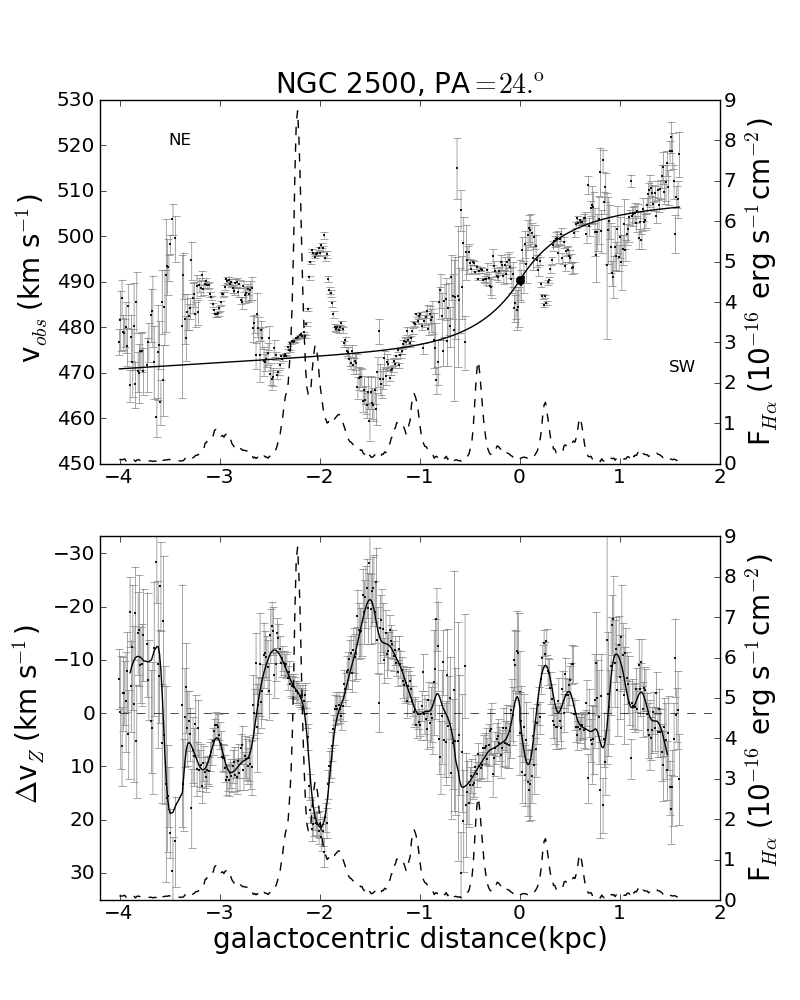}
\includegraphics[width=0.45\textwidth]{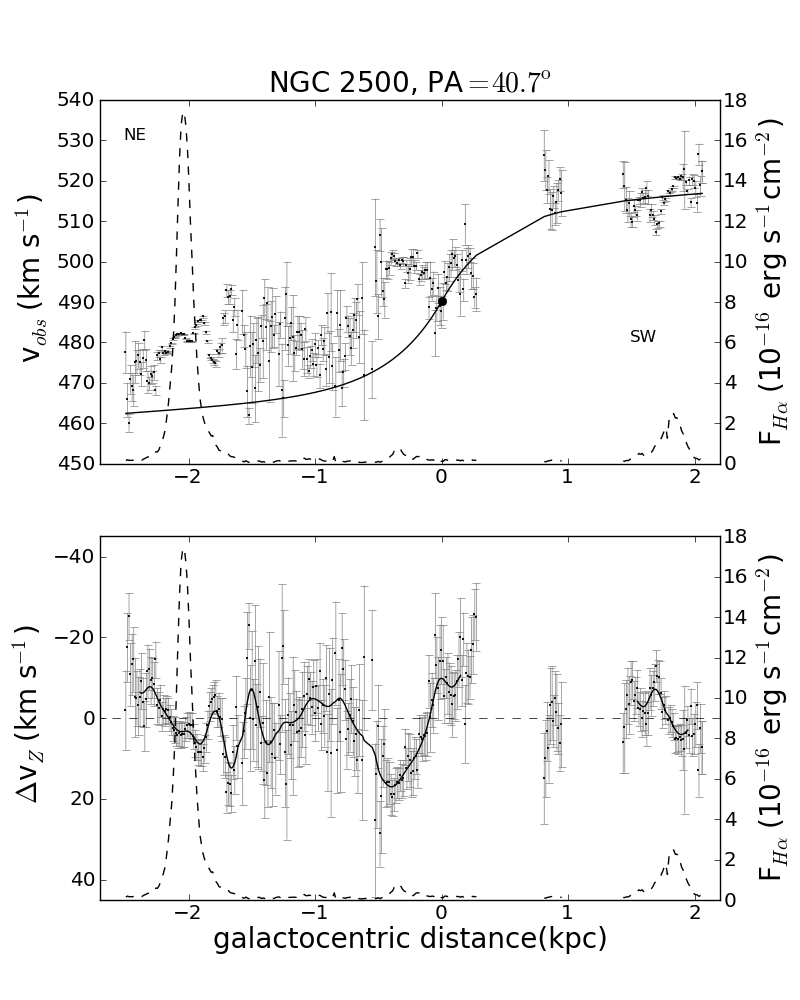}
\includegraphics[width=0.45\textwidth]{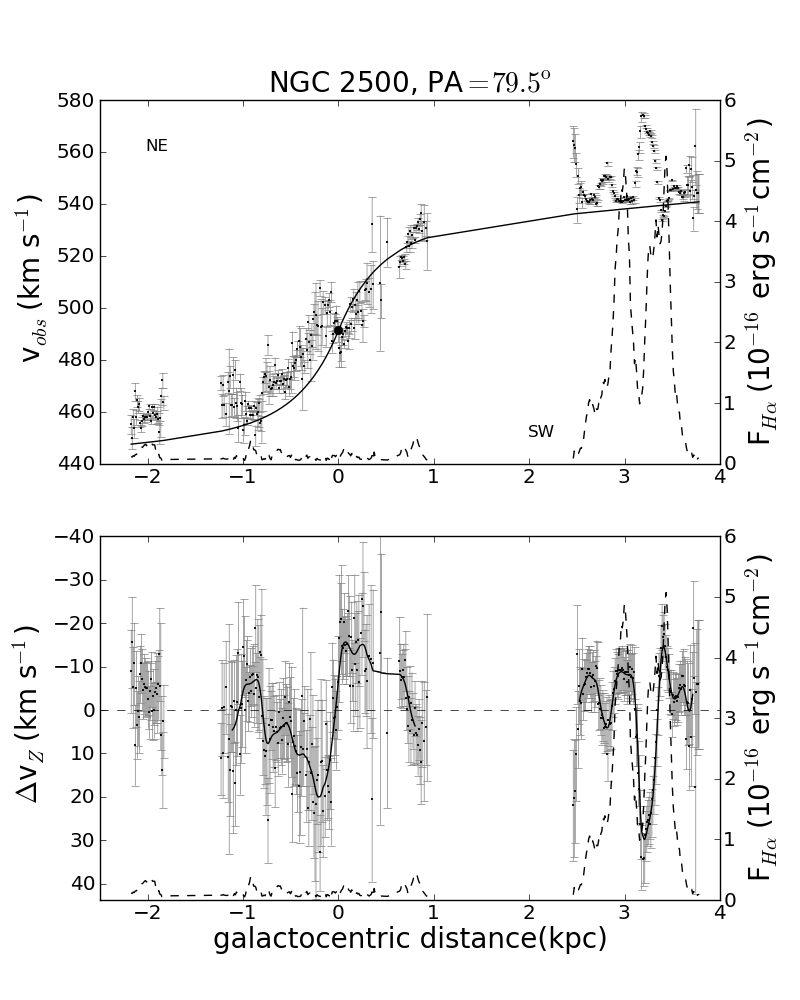}
\caption{ Top: The observed velocities for NGC~2500, derived from the \Ha\ emission lines (see Section \ref{SecObs}). 
The bold line corresponds { to} the projected rotation curve, and the black dot marks the systemic velocity. 
The \Ha\ intensity is drawn in the right-hand ordinate, with a dashed line. 
The {\it x-axis} represents the distance to the reference pixel, in kpc units;
Bottom: The detrended perpendicular velocity $\Delta V_Z$, at each pixel position across the slit, calculated as Equation \ref{eqVZ}. The black line is a smoothing line of $\Delta V_Z$, for an easier comparison with the \Ha\ emission lines.}
\label{fig4}
\end{figure*}

\begin{figure*}
\centering
\includegraphics[width=0.45\textwidth]{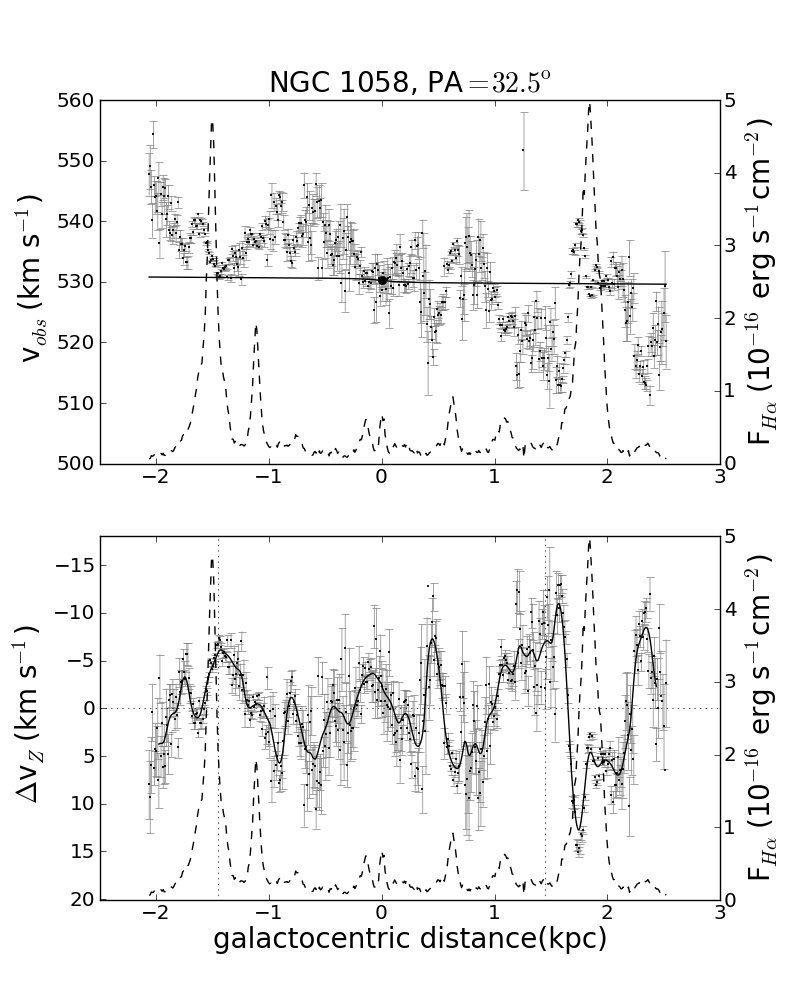}
\includegraphics[width=0.45\textwidth]{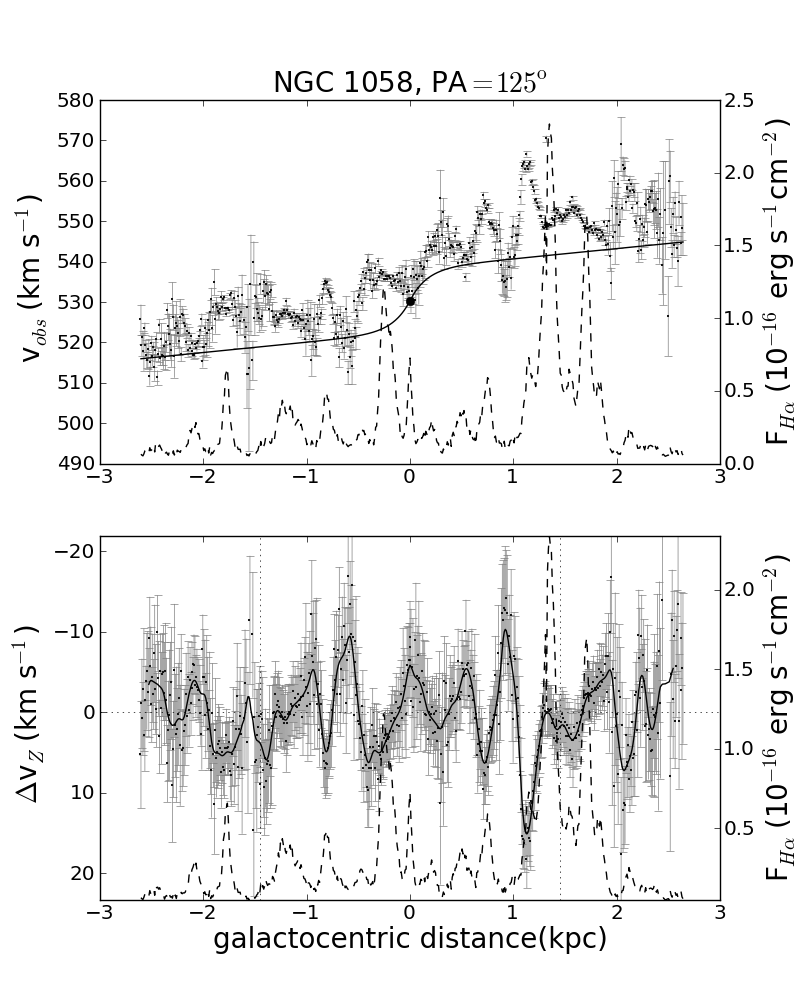}
\includegraphics[width=0.45\textwidth]{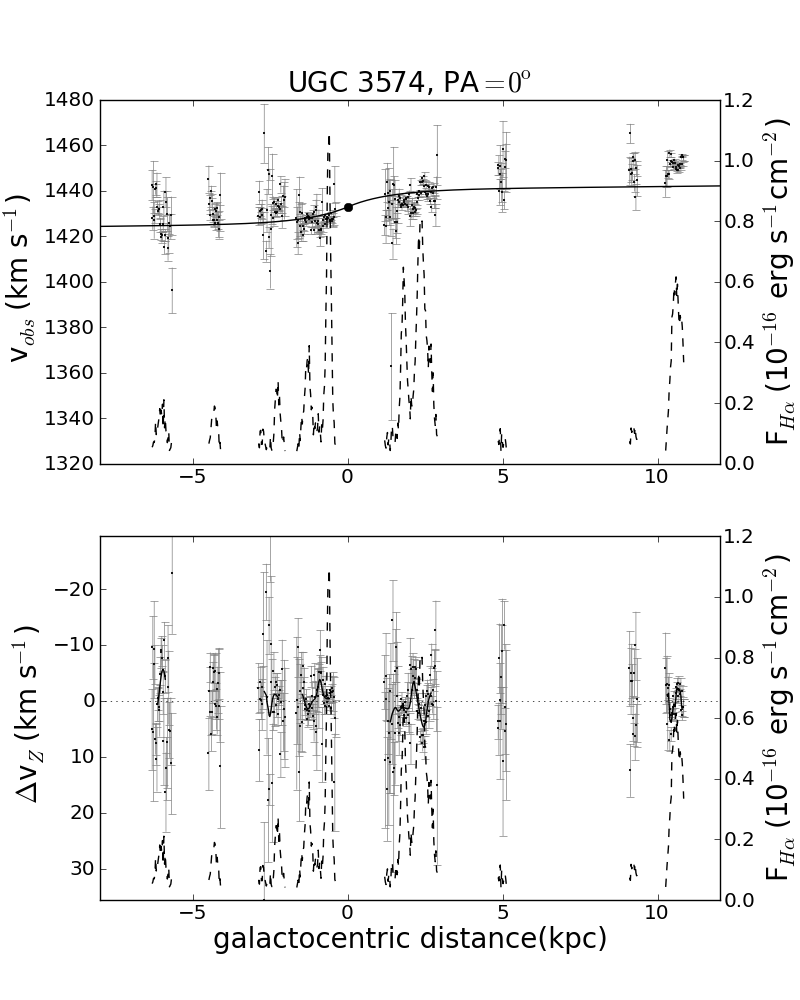}
\includegraphics[width=0.45\textwidth]{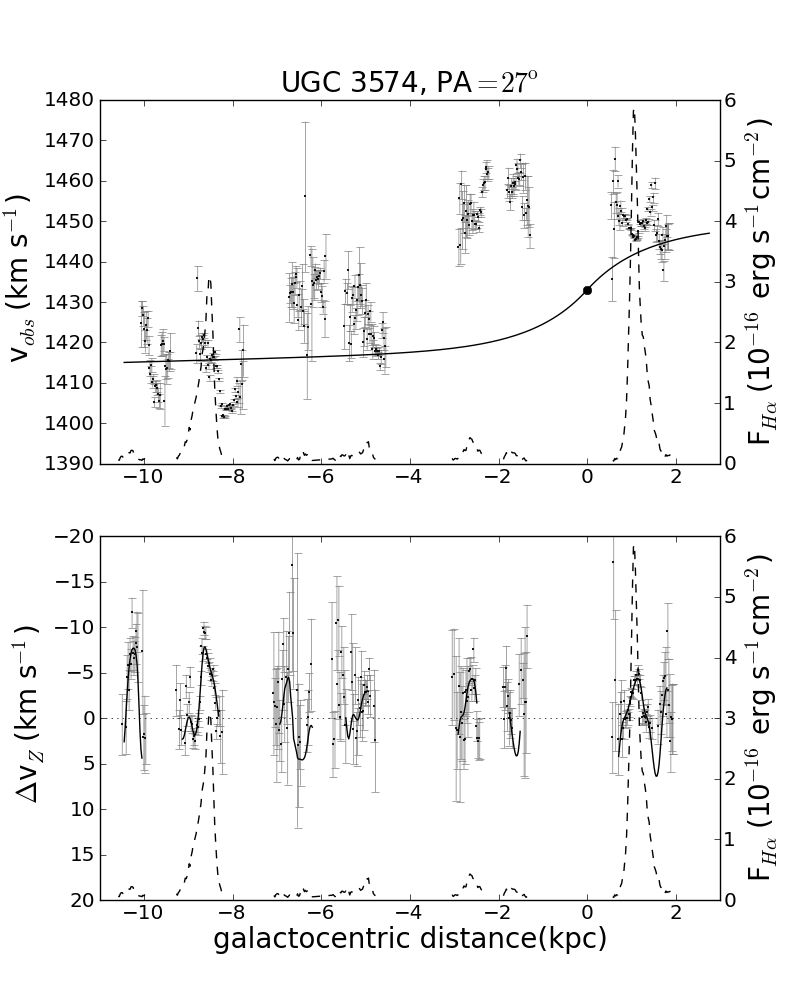}
\caption{ NGC~1058, PA = 32.5$^{\circ}$. 
Top: The observed velocities for NGC~1058 and UGC~3574, derived from the \Ha\ emission lines (see Section \ref{SecObs}). 
The bold line corresponds with the projected rotation curve, and the black dot marks the systemic velocity. 
This PA slit position is very close to the minor axis and the rotation curve projection appears to be rather constant, equal to the V$_{sys}$, since the rotational velocity is negligible. 
The \Ha\ intensity is drawn in the right-hand ordinate, with a dashed line. 
The {\it x-axis} represents the distance to the reference pixel, in kpc units;
Bottom: { The detrended perpendicular velocity by fitting a linear component to the global trend. }
The black line is a smoothing line of $\Delta V_Z$, for an easier comparison with the \Ha\ emission lines.}
\label{fig5}
\end{figure*}

The same procedure is applied for the rest of the galaxies, although, for the sake of clarity, only the observed velocities (on the top panels) and the final detrended vertical velocities (on the bottom) are shown in Figures \ref{fig4} and \ref{fig5}. 
In { the} next section we describe a more detailed analysis or description of the morphologies or patterns found from these plots for each galaxy.

\section{Results}\label{SecResults}

In this section we describe the deviations from the main rotation observed, $\Delta V_Z$, which we interpret as perpendicular motions of the ionized gas from the galactic disk, or corrugations.
The aim is to show the existence of such corrugations in galaxies, as well as giving a description of its morphologies or patterns. We found the maxima and minima values of $\Delta V_Z$ are usually related with \Ha\ emission line peaks, and with the morphology of the galaxy. 
Its interpretation is still an open matter of discussion.

For NGC~278 and NGC~1058 a wavelike structure of $\Delta V_Z$, along the whole slit, is clearer than for the other two galaxies. 
In NGC~2500 and UGC~3574 there are some regions with no \Ha\ emission or too faint to be fitted, and consequently there is no $\Delta V_Z$ associated with these regions. 
This last feature is more remarkable in UGC~3574, the most distant galaxy of the sample, { losing} therefore a global view of the vertical velocity deviations  with respect to the \Ha\ emission (Figure \ref{fig5}).

In a previous study for the corrugated velocities in NGC~5427, \citet{2001ApJ...550..253A} found a systematic spatial shift between the velocity peaks, $\Delta V_Z$, and the emission line intensity peaks of \Ha;
where the approaching (negative values) peaks of $\Delta V_Z$ occur in the convex border of the arms, and the receding (positive values) maxima are located behind the \Ha\ emission maxima, in the concave side. 
This kinematical behaviour is interpreted as the response of the gas flow into a spiral density wave in a thick and magnetized gaseous disk, described by  the \citet{1998ApJ...509..703M} and \citet{1999ApJ...526L..89M} galactic bore models. 

For each slit in all galaxies, we find some of these features of the variation of the vertical motions when the slit passes through spiral arms, but with a varied phenomenology. 

In NGC~278, the behaviour described by \citep{2001ApJ...550..253A} is clearly observed for the slit  PA = 40.5$^{\circ}$,  between galactocentric distances  $-1$ and $-0.5$ kpc, with $\Delta V_Z \sim 30$ \kms\ (Figure \ref{fig6}). 
This feature is in the concave side of the spiral arm, since it is located within the corotation radius ($\sim1.5$ kpc, \citealt{2009ApJS..182..559B}). The gas reaches the density wave behind the spiral arm border -or concave side-,  
so the approaching (negative values) peak in $\Delta V_Z$ occurs behind the \Ha\ emission maxima, or the corresponding spiral arm, when the gas is flowing up to higher galactic latitudes above the arm. Then it falls down onto the galactic plane, which corresponds to the receding maxima (positive values) in $\Delta V_Z$ that { appear ahead of }the arm. 

This phenomenon is also observable, although for fainter \Ha\ emission maxima, and close to the corotation radius, between galactocentric distances $-1.2$ and $-1.5$ kpc, with high vertical  $\Delta V_Z \sim 40$ \kms, and around $1$ kpc with $\Delta V_Z \sim 15$ \kms. 

Another interpretation for the approaching and receding peaks in $\Delta V_Z$ could be the deceleration and acceleration with respect to the rotation velocity. If this were the case, the observed velocities would be represented  by $v_{rot}$. 
If the vertical motions were neglected, the associated variation in the observed velocities of 
$\Delta V_{obs} \sim 20$ \kms\  cannot be explained by the difference between the rotation velocities. 

An even more outstanding example is found for the slit PA = 143.5$^{\circ}$, the minor axis. 
The most important peak in this case shows a variation in the observed velocities between  galactocentric distances  0.5 and 0.8 kpc. 
A $\Delta V_{obs}  \sim 22$ \kms\  cannot be explained by the difference between the rotation velocities  
$\Delta V_{rot} \cos \theta \sin i \sim 0.13$ \kms.
On the other hand, if the $\Delta V_Z$ were not considered and with a projection factor of $\cos \theta \sin i  \simeq -0.009$, for a $\Delta V_{obs} $ of $22$ \kms\ a difference for the rotation velocity  of the order of 2400 \kms\ would be required. 
This would imply a velocity radial gradient $\sim 8000$ \kms/\kpc{, which makes} no sense in a galactic disk and radial distances of 300 pc.

With our data and resolution we cannot discern if this is the case of a bubble in expansion or a merger, as the process that would explain such behaviour in $\Delta V_Z$. 
 Moreover, this feature at PA $= 143.5^{\circ}$ is found as the largest vertical displacement, whereas it does not seem to be related with a significant \Ha\ emission peak.
However, we cannot claim that it is not related with a spiral arm, since the weakness of \Ha\ emission does not determine the presence of a spiral arm.  

At slit PA$= 93.15^{\circ}$, between $\sim -1$ and $1.2$ kpc (close to corotation), we can identify a double $\Delta V_Z$ peak associated with \Ha\ emission peaks.
So its  interpretation in the framework of \citet{2001ApJ...550..253A} is not clear.
Although this is a quite plausible model, with the current data, we cannot discard other mechanisms as the origin or contributing to the enhancement of the corrugations in the velocity field. 

Similar patterns can be observed in the rest of  the galaxies. In some locations, it is possible to discern similar  behaviour to that expected from a galactic bore. 
For many others, the mechanism behind  it is not clear. However, the changes in $\Delta V_Z$ are still clearly seen, as it crosses a spiral arm (as identified by the \Ha\ peaks), from the receding to the approaching side, or vice versa. So these changes in $\Delta V_Z$ are definitely related to the spiral arms. 

Some examples of such patterns are found, for example in NGC~2500, between $-2.5$ and $-2$ kpc at PA$= 24^{\circ}$, $\Delta V_Z$ goes from $-10$ to $20$ \kms. Again, just after the spiral arm, the gas flows up to nearly $30$ \kms. 
At PA$= 40.7^{\circ}$, in the central part of the slit, $\Delta V_Z$ are more pronounced but they are also  noisier due to the faint \Ha. The stronger \Ha\ emission peaks are at the ends of the slit, where we can observe the pattern described above,  around $-2$ and $\sim1.8$ kpc, respectively. 
At PA$= 79.5^{\circ}$, close to the minor axis, this is observed between $3$ and $3.5$ kpc, where the gas flows up from $30$ to $-20$ \kms. Between the two arms there is a similar change in $\Delta V_Z$ from $-10$ to $30$ \kms.  As in the { previous} PA, there is a weaker and  noisier zone between $-2$ and $1$ kpc, but the corrugations are still observable. 

In NGC~1058, for PA$= 32.5^{\circ}$ and around the corotation radius at $-1.5$ kpc, there is another example of  $\Delta V_Z$ across the spiral arm. In the opposite side of the slit, between $1.5$ and $2$ kpc,  just before the spiral arm there is an important change of $\Delta V_Z=-10$ to $15$ \kms. 
Another example  of the pattern we are describing as similar to a galactic bore is  found at PA$= 125^{\circ}$, between $1$ and $1.5$ kpc, just before the arm (around $1$ kpc) { where} there is a large vertical deviation. At this PA the \Ha\ emission is bright enough that peaks have been  measured continuously, and  so  corrugations are more noticeable. 

UGC~3574 is the farthest galaxy of the sample, therefore the \Ha\ emission is fainter and more difficult to measure along the slit (Figure \ref{fig5}). Nevertheless, the vertical velocity deviations are appreciable and these seem to be linked to the spiral arms.  
At PA$= 27^{\circ}$, between $-9$ to $-8$ kpc and $1$ to $2$ kpc we can see how the $\Delta V_Z$ changes across the spiral arms.

In summary, we found in some cases that it is easy to see how  $\Delta V_Z$ changes across the arm, quite similar to the pattern described by \citet{2001ApJ...550..253A}, and it could be explained by galactic bore models. 
But in many other cases,  $\Delta V_Z$ changes are shifted, or they become wider, or lengthened beyond the arm. 
That is, there are other features where the expected pattern for this model is not so clear or { is not} observed. 
In some cases, an important vertical displacement $\Delta V_Z$ does not seem { to be} related with an \Ha\ emission peak, as for example the feature described above at PA $= 143.5^{\circ}$ in NGC~278 or between $-1$ and $0$ \kpc\ at PA $= 40.7^{\circ}$ in NGC~2500.  

\subsection{Comparison between Observations and Models}

Our analysis establishes the existence of a vertical velocity field that shows a sinusoidal structure associated mainly with the observed spiral arms. How can the generation of these structures be explained? Since the 1970s, different models have been proposed to explain the genesis of the spatial corrugations observed in our Galaxy and other external galaxies (see \citet{A96} for a review of this topic to that date). But, regarding the variations of the vertical velocity field, the first article that took the possible corrugation of velocities into consideration was published by \citet{2001ApJ...550..253A}, for the galaxy NGC~5427, in which the hydraulic bore mechanism  \citep{1998ApJ...509..703M,2004ApJ...615..758G,2004ApJ...615..744G} was proposed for the formation of the velocity field observed in the vicinity of the spiral arms. The choice of this model, for comparison with the observations of the velocity field of NGC 5427, was determined by it being the only model that predicted the expected velocity structure in the neighbourhood of spiral arms, and therefore the only one that made it possible to perform the comparison. Moreover, both the quantitative and qualitative aspects of the model fitted with the velocity field observed. The situation has not changed much over the last decade, and few models of the Galaxy, if any, incorporate corrugations into their description. Yet, what is the principal ingredient that makes the models proposed by the Wisconsin group (and successors) different to the rest? The answer is simple: the magnetic field. The majority of models prior to the one proposed by \citet{1998ApJ...509..703M} are based on the interaction between different galactic subsystems (like discs and halos), and the impact of the disc with high-velocity clouds (whether HI or dark matter), and do not include the magnetic field in their physical fundamentals -- with the exception of \citet{1999ApJ...515..657S} on the impact with HVCs, and those on the Parker instability for the generation of super-clouds and corrugations in spiral arms \citep[i.e.][]{2002ApJ...570..647F}. Should we include the magnetic field in the physical foundations of any model that tries to explain the observed corrugations? Our answer is yes, and there are two types of arguments that back this affirmation: a) there is a non-negligible magnetic field in our Galaxy \citep{2015A&A...576A.104P,2015ASSL..407..507B}; and b) in the evolutionary models of the Galactic interstellar medium, the inclusion of the Galactic magnetic field is far from being irrelevant \citep{1998ApJ...509..703M,1999ApJ...515..657S,2004ApJ...615..758G,2004ApJ...615..744G}.

\section{Diagnostic Diagrams}\label{SecDD}

To complement the kinematic analysis, diagnostic diagrams (DD) are included to study the ionization structure in the observed features. 
As { it is }well known, the ionization and excitation process can be determined by plots of [NII]/\Ha\ versus [OIII]/H$\beta$, depending on its location in such diagrams \citep{1981PASP...93....5B}. These ratios are the best options among all the possible emission line ratios \citep{1987ApJS...63..295V}, and therefore one of the most commonly and widely used DD. 

Before plotting the DD, the flux ratios are corrected for reddening following the relation 

\begin{equation}
\frac{I_{H\alpha}}{I_{H\beta}} = \frac{F_{H\alpha}}{F_{H\beta}} \times 10^{0.4(A_{H\alpha} - A_{H\beta})}= 
\frac{F_{H\alpha}}{F_{H\beta}} \times 10^{0.4 A_V (k_{H\alpha} - k_{H\beta})}
\end{equation}

\noindent where $I_{\lambda}$ is the unreddened emission at a certain wavelength from the source, and $F_{\lambda}$ is the observed flux, affected by dust absorption. $A_{\lambda} = k_{\lambda}A_V$ is the extinction factor at a particular waveband $\lambda$, and $k_{\lambda}$ and $A_V$ are the extinction curve and the absorption in the V band, respectively. 
The Balmer decrement is a { method} for reddening correction, deduced from the comparison of the observed \Ha\ to H$\beta$ flux ratio with the theoretical value of $2.86$ \citep{1989agna.book.....O}, based on case B recombination. Applying the Balmer decrement we have that,

\begin{equation}
A_V = \frac{log(2.86) - log(F_{H\beta} / F_{H\alpha})}{0.4(k_{H\alpha} - k_{H\beta})}
\end{equation}

So, the unreddened flux ratios can be obtained as

\begin{eqnarray}
log\Big(\frac{I_{\lambda1}}{I_{\lambda2}}  \Big) &=& log\Big(\frac{F_{\lambda1}}{F_{\lambda2}}  \Big) + 0.4A_V(k_{\lambda1} - k_{\lambda2})=\nonumber\\
&=& log\Big(\frac{F_{\lambda1}}{F_{\lambda2}}  \Big) + C_1 log\Big(\frac{F_{H\alpha}}{F_{H\beta}} \Big) - C_2
\end{eqnarray}

\noindent where $C_1 = \frac{k_{\lambda2} - k_{\lambda1}}{k_{H\alpha} - k_{H\beta}}$, and $C_2 = log(2.86)C_1$.

\begin{figure*}
\includegraphics[height =0.49\textwidth]{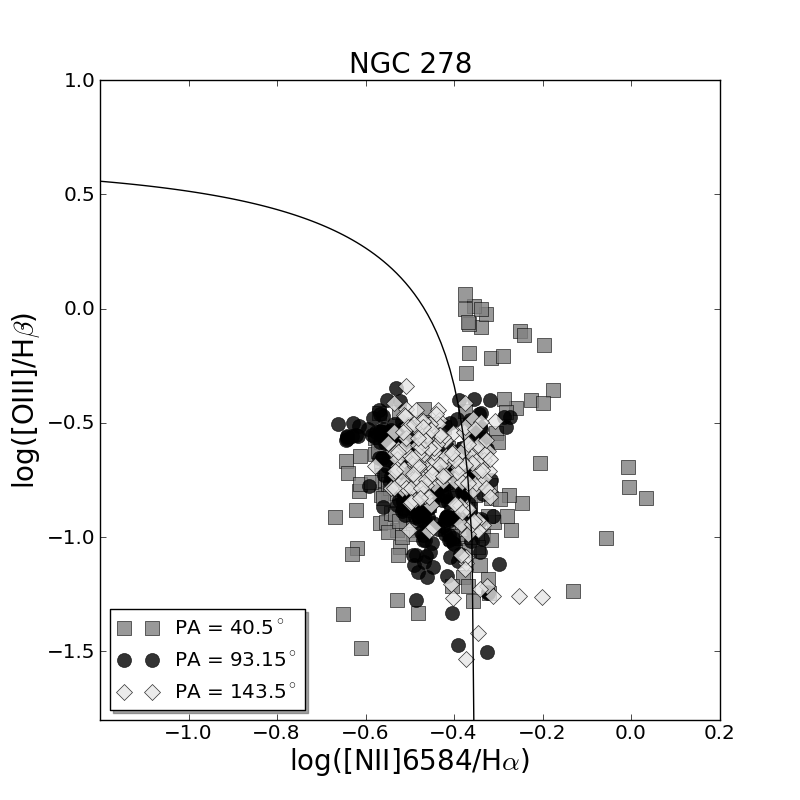}
\includegraphics[width=0.49\textwidth]{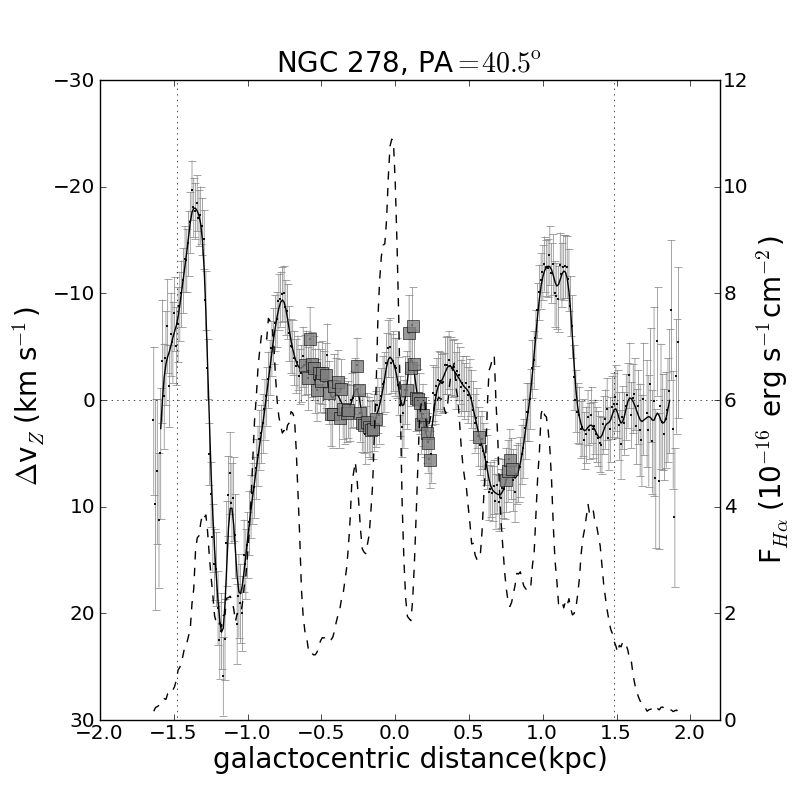}
\includegraphics[width=0.49\textwidth]{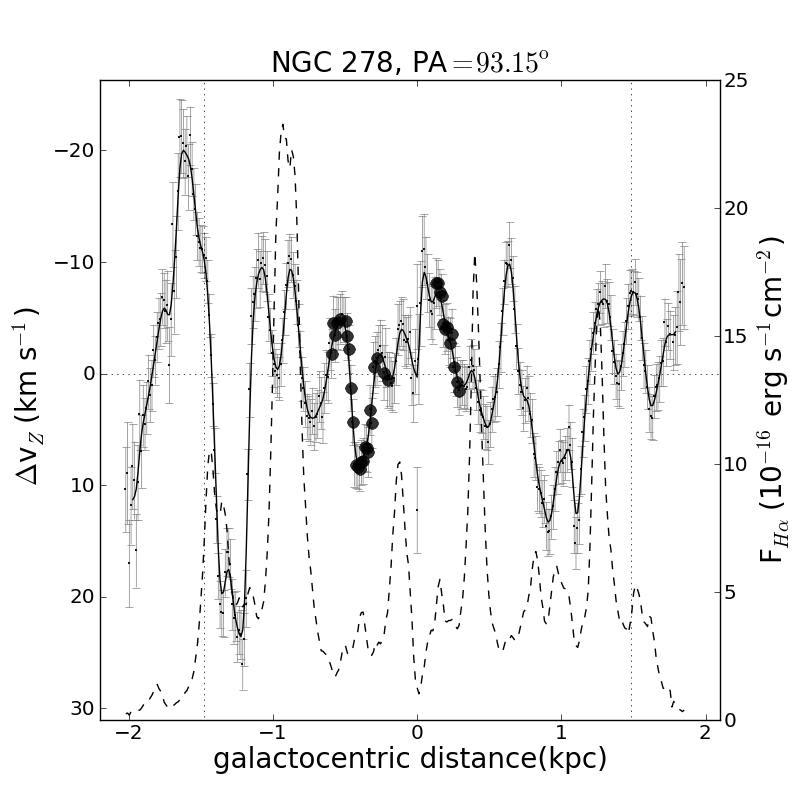}
\includegraphics[width=0.49\textwidth]{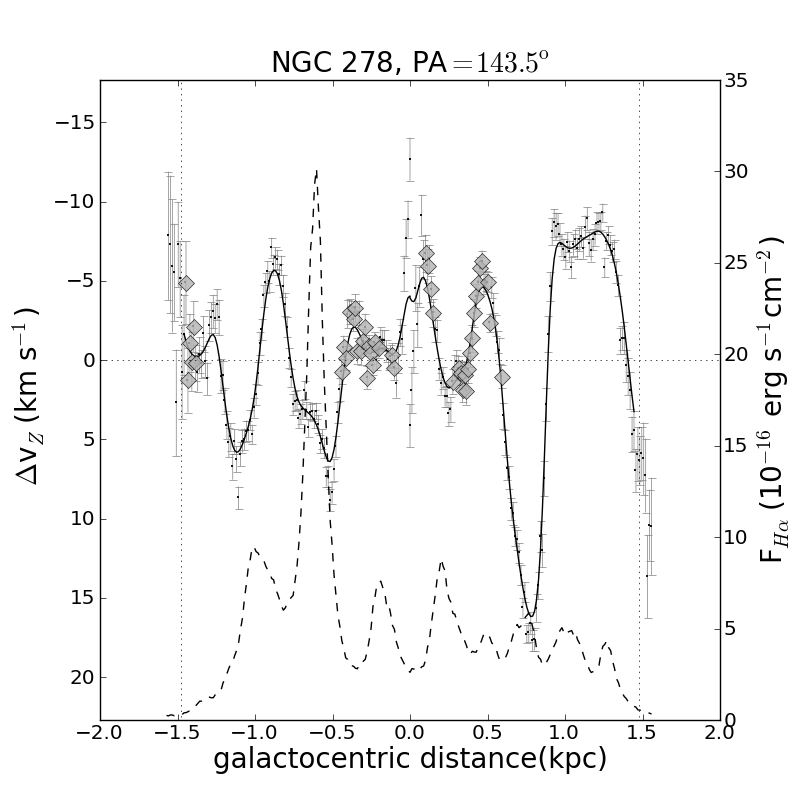}
\caption{  
Top left: Diagnostic Diagram [NII]/\Ha\ vs. [OIII]/H$\beta$ \citep{1987ApJS...63..295V}, for NGC~278.
The resting plots are the detrended perpendicular velocity $\Delta V_Z$ for the different slit positions. In this case, only those  pixels with error $\delta V_Z < \overline{\delta V_Z}$ are plotted. 
The highlighted pixels correspond to those locations where the ionization mechanism is not photoionization. 
The \Ha\ intensity is drawn in the right-hand ordinate, with a dashed line. 
A dotted line marking the zero value { stands out} as the global trend has been practically removed, 
 with only the local oscillations of $V_Z$ remaining. The {\it x-axis} represents the distance to the reference pixel, in kpc units.}
\label{fig6}
\end{figure*}

\begin{figure*}
\includegraphics[height =0.49\textwidth]{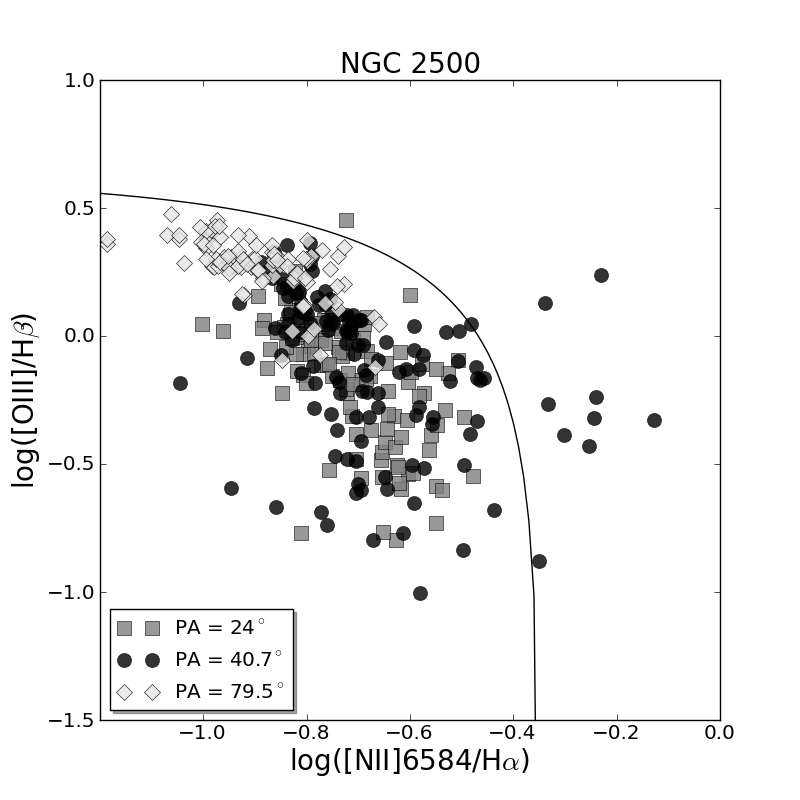}
\includegraphics[width=0.49\textwidth]{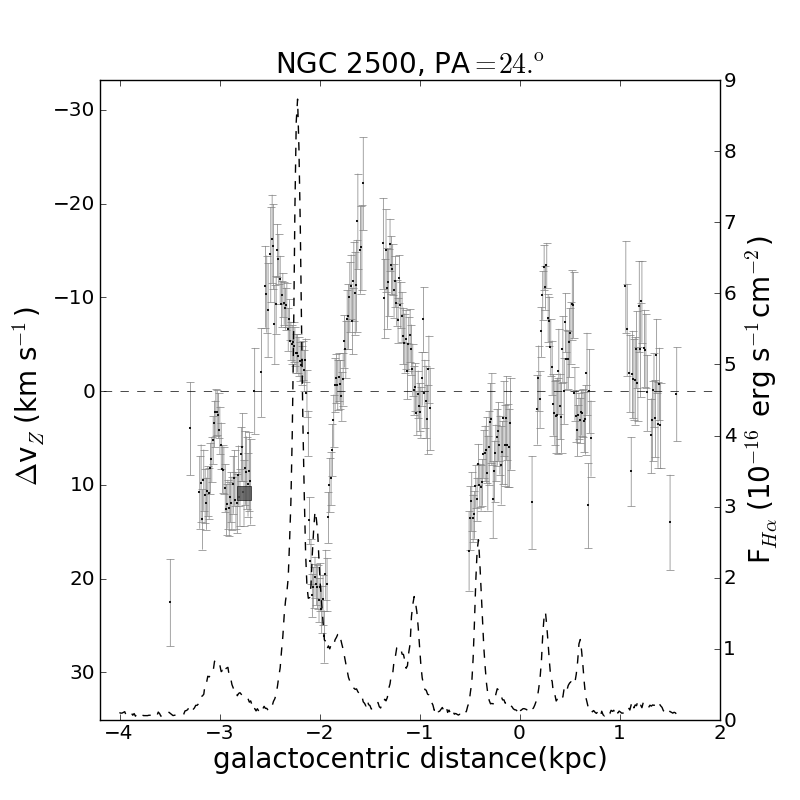}
\includegraphics[width=0.49\textwidth]{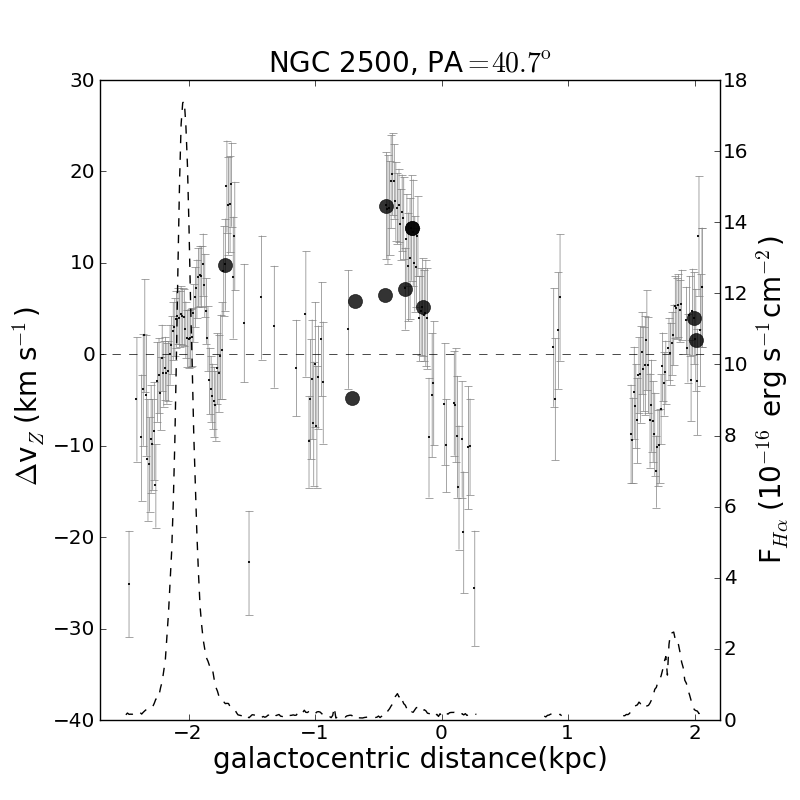}
\includegraphics[width=0.49\textwidth]{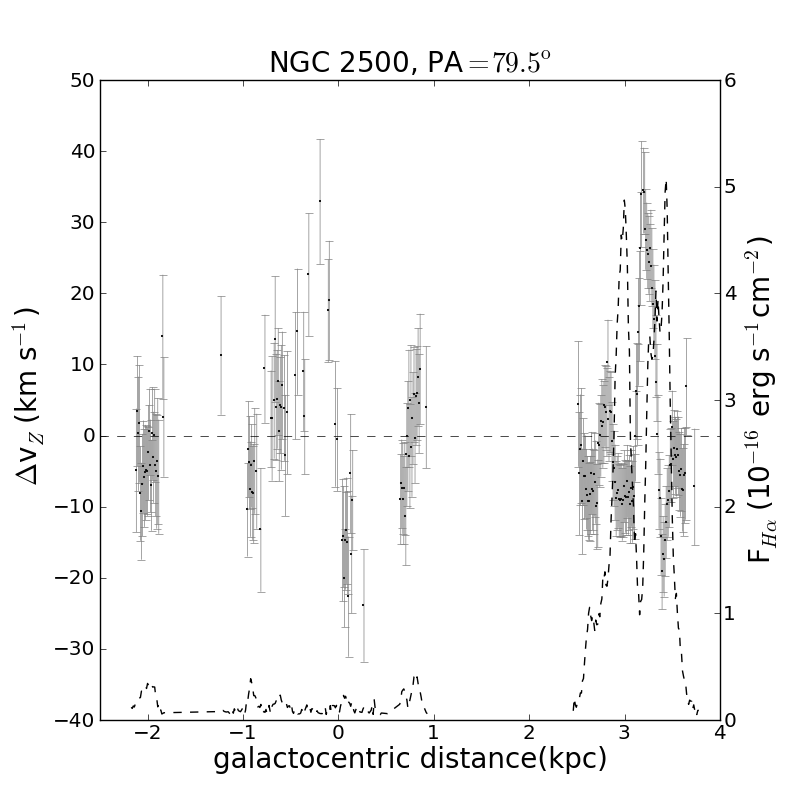}
\caption{ Analogous to Figure \ref{fig6}, but for NGC~2500.}
\label{fig7}
\end{figure*}

\begin{figure*}
\includegraphics[height =0.49\textwidth]{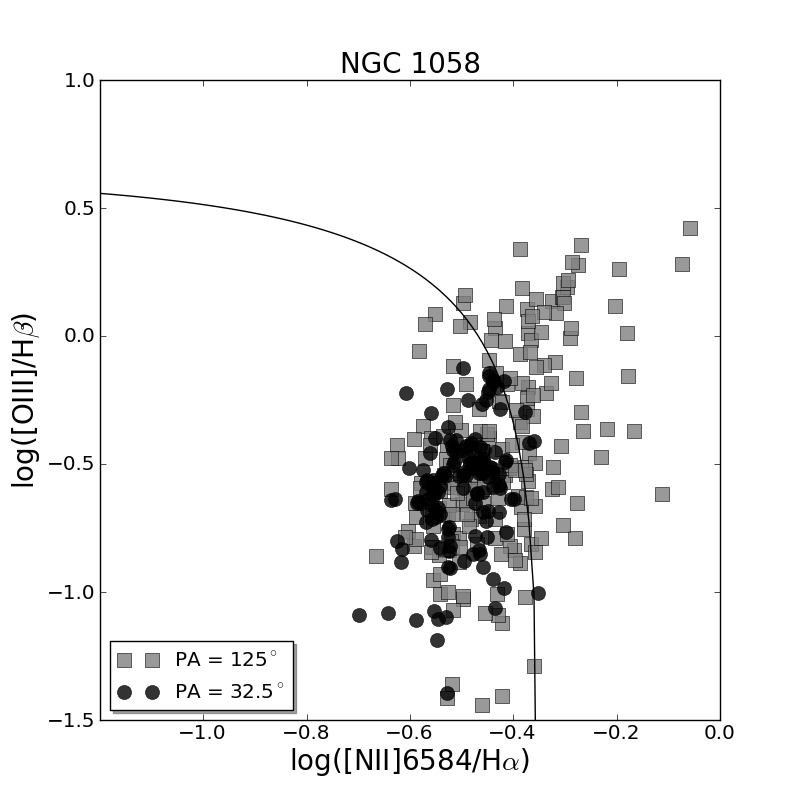}
\includegraphics[width=0.49\textwidth]{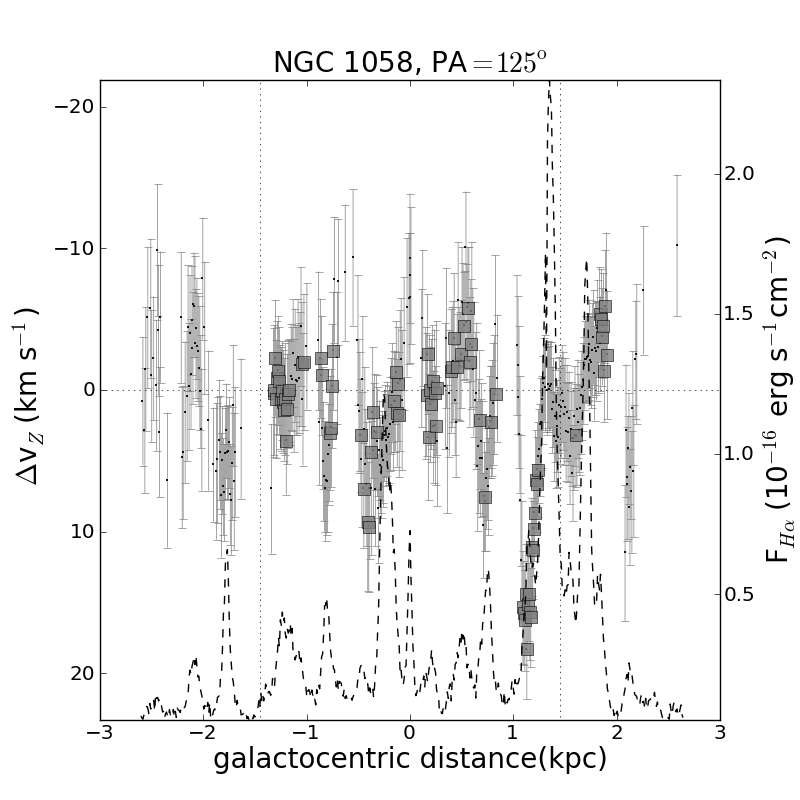}
\includegraphics[width=0.49\textwidth]{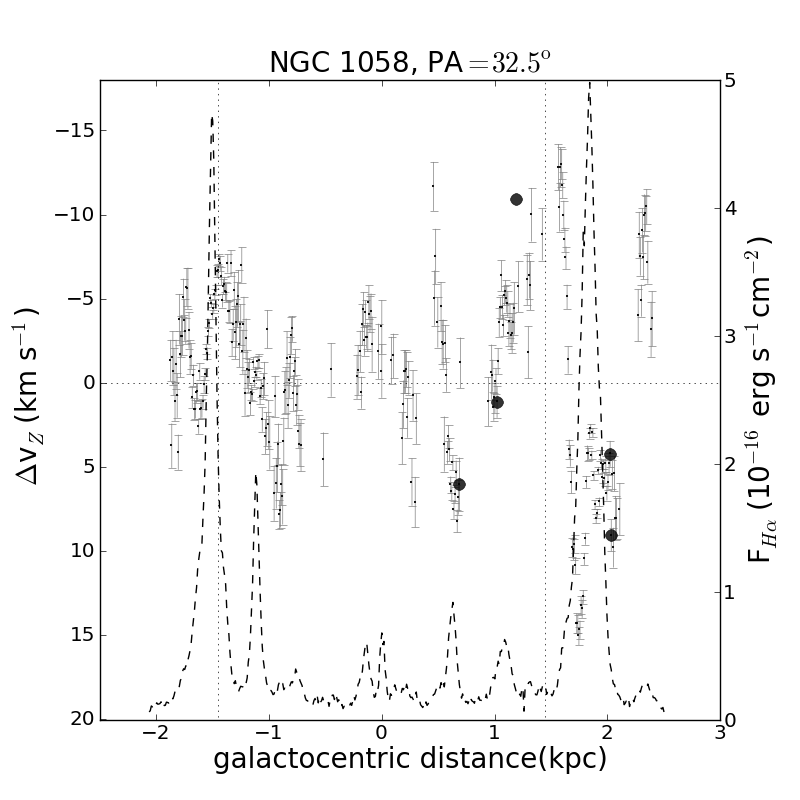}
\caption{ Analogous to Figure \ref{fig6}, but for NGC~1058.}
\label{fig8}
\end{figure*}

\begin{figure*}
\includegraphics[height =0.49\textwidth]{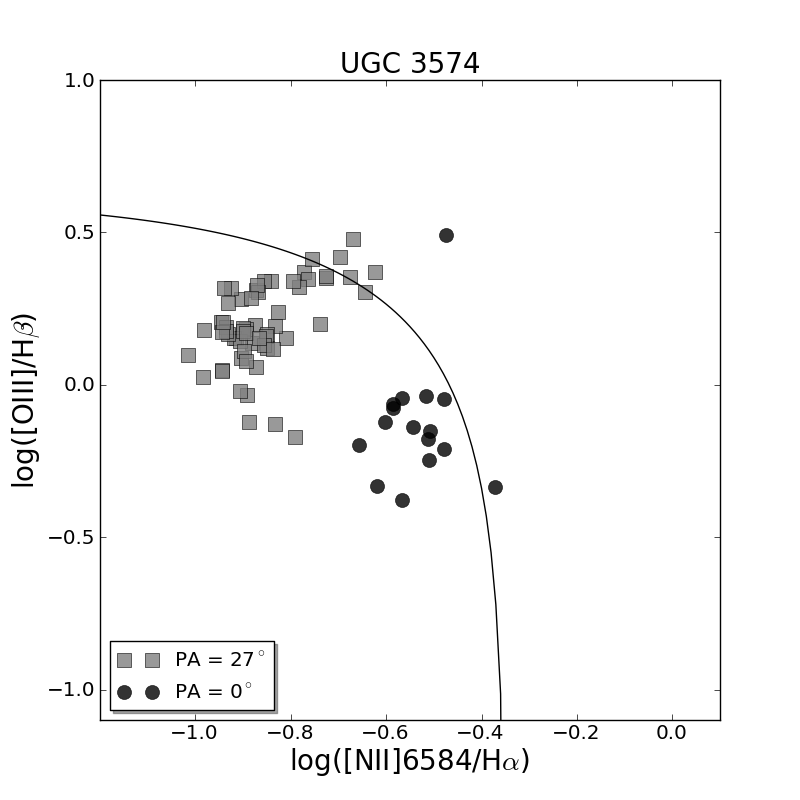}
\includegraphics[width=0.49\textwidth]{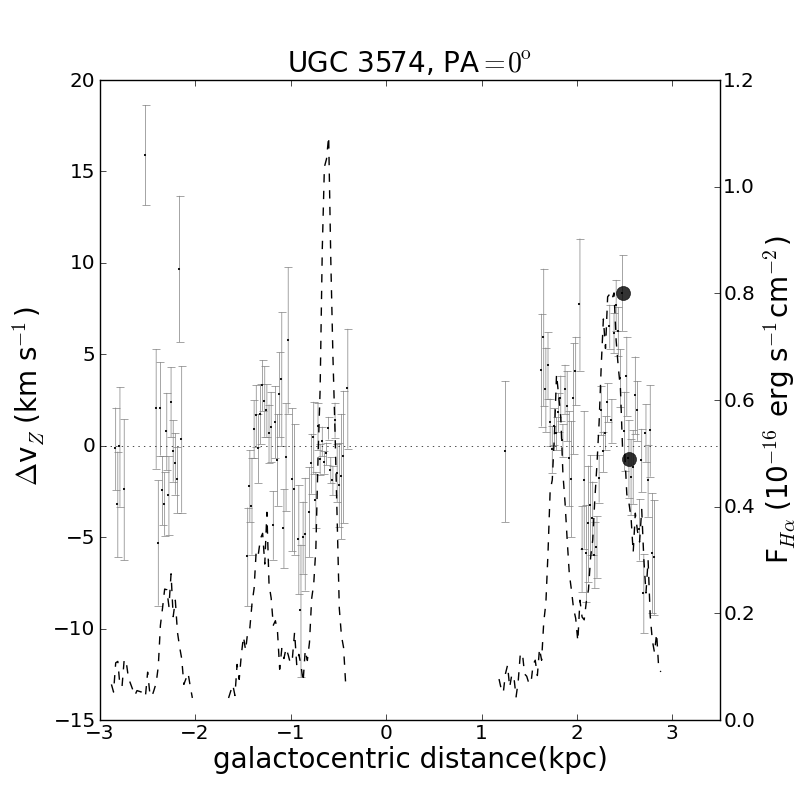}
\includegraphics[width=0.49\textwidth]{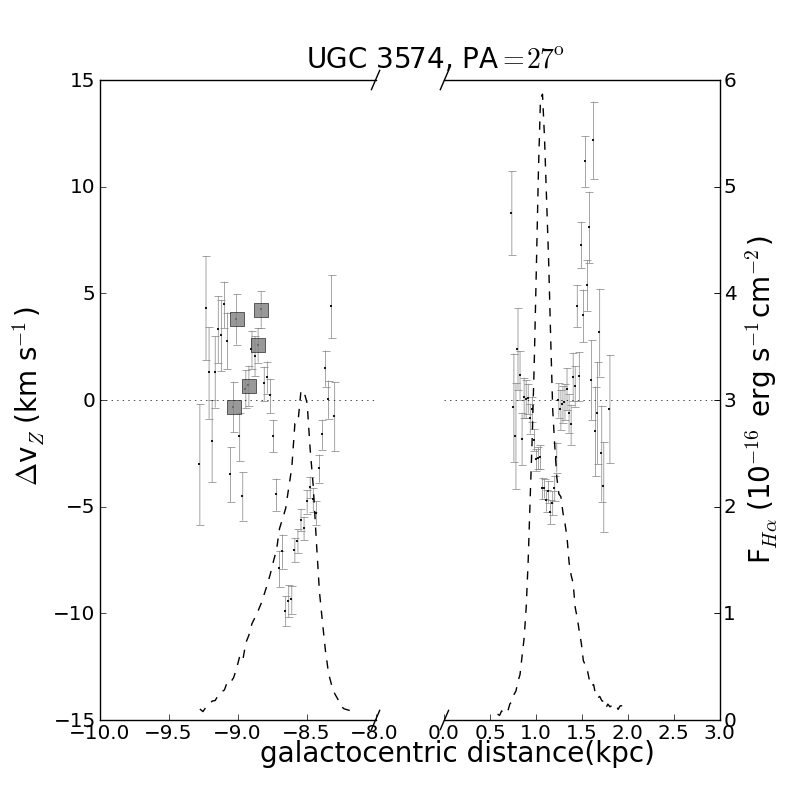}
\caption{ Analogous to Figure \ref{fig6}, but for UGC~3574.}
\label{fig9}
\end{figure*}

The DD for each galaxy are found in Figures \ref{fig6} to \ref{fig9}, together with the plots of $\Delta V_{Z}$. Plotted individually, they have a better resolution than the above plots of $\Delta V_{Z}$ in Figures \ref{fig3}$-$\ref{fig5}. 
To include more robust information, only those  pixels with $\Delta V_Z$ less than the mean of the errors $\overline{\Delta V_Z}$ are plotted. 
So they  clearly show the relationship between the velocity and the \Ha\ emission line flux peaks. 

The different slit position angles for each galaxy are represented with different symbols and  { grey colour} scale, as indicated in the figures. The locations where the DD ratios are above the reference line for photoionization by young stars are shown in the panels with the $\Delta V_{Z}$ kinematics. 
This could provide some clues as to whether the ionization mechanism is related to the presence of  corrugations and the encounter of the density wave with a spiral arm. 
Within the galactic bore model framework \citep{1998ApJ...509..703M,1999ApJ...526L..89M}, we would expect the presence of shocks  at these locations. 

No concluding results are obtained from the DD,  which show that photoionization by young massive stars seems to be the main ionization mechanism in most locations. 

There is only a small portion of the gas that appears to be ionized by low-velocity shocks. In NGC~278 and in NGC~1058, at PA$=125^{\circ}$, this portion is more important. 
In NGC~278, for example, these low-velocity shocks would be  located in the innermost central \kpc, coinciding with the minor merger found by \citet{2004A&A...423..481K}.

Moreover, two different  forms of behaviour appear in the [NII]/\Ha\ {\it vs.} [OIII]/H$\beta$ diagnostic diagrams. 
NGC~278 and NGC~1058 have most of their pixels, for the three slit positions, concentrated at the same location in the DD, whereas for NGC~2500 and UGC~3574 different slit positions occupy a different location in the DD, nearly covering the theoretical curve separating the different ionization mechanisms. 
This  behaviour may be related to a metallicity gradient, but a more detailed study,  beyond the scope of this paper, would be needed. 

In summary, although the association between the gas ionization mechanisms and kinematic properties do not offer a conclusive pattern, some general characteristics are observed: a) most of the gas is ionized by high-energy photons; and b) in NGC 278 and NGC 1058 there is a not negligible number of points whose diagnostic diagram suggests ionization through low-velocity shocks. In NGC 278 these shocks seem particularly limited to the central region where some authors have observed signs of an inner merger \citep{2004A&A...423..481K}. Otherwise,  we note the potential to explore this idea further with the integral field spectroscopic data sets that are becoming available, such as from the surveys CALIFA \citep{2012A&A...538A...8S,2015A&A...576A.135G}, MaNGA \citep{2015ApJ...798....7B}, and SAMI \citep{2012MNRAS.421..872C}, and with  much better resolution from data already becoming available from the MUSE instrument at the VLT \citep{2010SPIE.7735E..08B}.

\section{Summary} \label{SecSum}

In this work we extend the detection and study of kinematic corrugations, already reported in the literature \citep[e.g.][]{2001ApJ...550..253A}, to a small sample of face on spirals.

The galaxies studied here show a similar  behaviour to NGC~5427 \citep{2001ApJ...550..253A} at some locations, with a clear displacement between the velocity and the emission line peaks, where the approaching peaks of $\Delta V_Z$ occur in the convex side of the spiral arms, and the receding maxima are located behind the \Ha\ flux maxima, in the concave side. This kinematical  pattern is similar to the one expected in a galactic bore generated by the interaction of a spiral density wave with a thick magnetized gaseous disk.  

In locations where $\Delta V_Z$ peaks are found to be related to \Ha\ emission maxima, a strong velocity gradient crossing the arm is sometimes observed, as seen in Figures \ref{fig6}  to \ref{fig9}, and described in Section \ref{SecResults}.  

This relation is not so clear in NGC~2500 and UGC~3574.  However, in these galaxies the \Ha\ emission is much fainter, and the \Ha\ emission line could only be fitted in a limited number of points along the slit. This produces a bias, and no clear conclusions can be deduced for these two galaxies.  

In NGC~278, at PA=143.5$^{\circ}$  there is a strong $\Delta V_Z$ peak in a weak \Ha\ emission zone. This could be associated with corotation or with a recent merger in the inner disk. We { have} checked that the stronger velocity peaks do not  necessarily coincide with the stronger emission line intensity peaks.

From the analysis of the ionized emission line diagnostic diagrams, we conclude that  photoionization by young stars is the main mechanism, and only a small portion of the gas  appear to be ionized by alternative mechanisms, such as low velocity shocks.
In this DD we see different behaviour for the two galaxies NGC~278 and NGC~1058 on one hand, and NGC~2500 and UGC~3574 on the other hand. The former have most of the pixels concentrated at the same location of the DD, whereas for the latter they are distributed covering more of the parameter space along the theoretical curve separating the different ionization mechanisms. This may be understood as a consequence of metallicity gradients, but further investigations are warranted. 
 
The origin of kinematic corrugations is still a matter of debate. Corrugations are closely linked, as cause-effect, to the large scale star formation processes: density waves, tidal interactions, collisions of high velocity clouds with disk, or a galactic bore generated by the interaction of a spiral density wave with a thick gaseous disk, etc. Which mechanism is the origin of disk gas corrugations is still an open question.


\section*{acknowledgment}

{ 
We are deeply grateful to the referee for constructive comments and valuable advice.
We acknowledge financial support from { the} Spanish Ministry of Economy and Competitiveness through grants AYA2010-17631 and AYA2013-40611-P, and from { the} Consejer\'{\i}a de Educaci\'on y Ciencia (Junta de Andaluc\'{\i}a) through TIC-101, TIC-4075 and TIC-114.}


\appendix
\bsp

\label{lastpage}

\end{document}